\documentclass{emulateapj}
\usepackage{amsmath}
\usepackage{graphicx}
\usepackage{natbib}
\usepackage[scaled]{helvet}
\usepackage{epsfig}
\usepackage{url}
\bibpunct{(}{)}{;}{a}{}{,}
\newcommand{\kms}{\,km\,s$^{-1}$}
\usepackage{textcomp}
\usepackage{mathpazo}
\usepackage{xspace}
\usepackage{apjfonts}
\usepackage{rotating}

\usepackage{color}
\usepackage[normalem]{ulem}

\newcommand{\bjdtdb}{\ensuremath{\rm {BJD_{TDB}}}}
\newcommand{\feh}{\ensuremath{\left[{\rm Fe}/{\rm H}\right]}}

\newcommand{\teff}{\ensuremath{T_{\rm eff}}\xspace}
\newcommand{\logg}{\ensuremath{\log g}}

\newcommand{\msun}{\ensuremath{\,M_\Sun}}
\newcommand{\rsun}{\ensuremath{\,R_\Sun}}
\newcommand{\lsun}{\ensuremath{\,L_\Sun}}

\newcommand{\rearth}{\ensuremath{\,R_{\rm \Earth}}\xspace}

\newcommand{\re}{\ensuremath{\,R_{\rm \Earth}}\xspace}
\newcommand{\me}{\ensuremath{\,M_{\rm \Earth}}\xspace}
\newcommand{\fave}{\langle F \rangle}
\newcommand{\fluxcgs}{10$^9$ erg s$^{-1}$ cm$^{-2}$}

\newcommand{\Kepler}{{\it Kepler}}
\newcommand{\Ktwo}{{\it K2}}
\newcommand{\ms}{\,m\,s$^{-1}$}
\newcommand{\thisstar}{K2-266\xspace}
\newcommand{\thisstarcomp}{EPIC248435395\xspace}

\newcommand{\vespa}{{\texttt{vespa}}\xspace}

\begin{document}

\title{A Compact Multi-Planet System With A Significantly Misaligned Ultra Short Period Planet}
\author{Joseph E. Rodriguez$^1$, Juliette C. Becker$^{2}$, Jason D. Eastman$^{1}$, Sam Hadden$^1$, Andrew Vanderburg$^{3,\star}$, Tali Khain$^{4}$, Samuel N. Quinn$^{1}$, Andrew Mayo$^{5,6,7,\dagger,\ddagger}$, Courtney D. Dressing$^{5}$, Joshua E. Schlieder$^{8}$, David R. Ciardi$^{9}$, David W. Latham$^{1}$, Saul Rappaport$^{10}$, Fred C. Adams$^{2,4}$, Perry Berlind$^1$, Allyson Bieryla$^1$, Michael L. Calkins$^1$, Gilbert A. Esquerdo$^1$, Martti H. Kristiansen$^{5,11}$, Mark Omohundro$^{12}$, Hans Martin Schwengeler$^{12}$, Keivan G. Stassun$^{13,14}$,\\ and Ivan Terentev$^{12}$}

\affil{$^{1}$Harvard-Smithsonian Center for Astrophysics, 60 Garden St, Cambridge, MA 02138, USA}
\affil{$^{2}$Astronomy Department, University of Michigan, 1085 S University Avenue, Ann Arbor, MI 48109, USA}
\affil{$^{3}$Department of Astronomy, The University of Texas at Austin, Austin, TX 78712, USA}
\affil{$^{4}$Physics Department, University of Michigan, Ann Arbor, MI 48109, USA}
\affil{$^{5}$Astronomy Department, University of California Berkeley, Berkeley, CA 94720-3411, USA}

\affil{$^{6}$DTU Space, National Space Institute, Technical University of Denmark, Elektrovej 327, DK-2800 Lyngby, Denmark}
\affil{$^{7}$Centre for Star and Planet Formation, Niels Bohr Institute \& Natural History Museum, University of Copenhagen, DK-1350 Copenhagen, Denmark}
\affil{$^{8}$Exoplanets \& Stellar Astrophysics Laboratory, Code 667, NASA Goddard Space Flight Center, Greenbelt, MD, USA}
\affil{$^{9}$NASA Exoplanet Science Institute, California Institute of Technology, Pasadena, CA, USA}
\affil{$^{10}$Department and Kavli Institute for Astrophysics and Space Research, Massachusetts Institute of Technology, Cambridge, MA 02139, USA}
\affil{$^{11}$Brorfelde Observatory, Observator Gyldenkernes Vej 7, DK-4340 T\o{}ll\o{}se, Denmark}
\affil{$^{12}$Citizen Scientist}
\affil{$^{13}$Department of Physics and Astronomy, Vanderbilt University, Nashville, TN 37235, USA}
\affil{$^{14}$Department of Physics, Fisk University, Nashville, TN 37208, USA}
\affil{$^{\star}$NASA Sagan Fellow}
\affil{$^{\dagger}$National Science Foundation Graduate Research Fellow}
\affil{$^{\ddagger}$Fulbright Fellow}

\shorttitle{}
\shortauthors{Rodriguez et al.}

\begin{abstract}
We report the discovery of a compact multi-planet system orbiting the relatively nearby (78pc) and bright ($K=8.9$) K-star, K2-266 (EPIC248435473). We identify up to six possible planets orbiting K2-266 with estimated periods of P$_b$ = 0.66, P$_{.02}$ = 6.1, P$_c$ = 7.8, P$_d$ = 14.7, P$_e$ = 19.5, and P$_{.06}$ = 56.7 days and radii of R$_P$ = 3.3 R$_{\oplus}$, 0.646 R$_{\oplus}$, 0.705 R$_{\oplus}$, 2.93 R$_{\oplus}$, 2.73 R$_{\oplus}$, and 0.90 R$_{\oplus}$, respectively. We are able to confirm the planetary nature of two of these planets (d \& e) from analyzing their transit timing variations ($m_d= 8.9_{-3.8}^{+5.7} M_\oplus$ and $m_e=14.3_{-5.0}^{+6.4} M_\oplus$), confidently validate the planetary nature of two other planets (b \& c), and classify the last two as planetary candidates (K2-266.02 \& .06). From a simultaneous fit of all 6 possible planets, we find that K2-266 b's orbit has an inclination of 75.32$^{\circ}$ while the other five planets have inclinations of 87--90$^{\circ}$. This observed mutual misalignment may indicate that K2-266 b formed differently from the other planets in the system. The brightness of the host star and the relatively large size of the sub-Neptune sized planets d and e make them well-suited for atmospheric characterization efforts with facilities like the Hubble Space Telescope and upcoming James Webb Space Telescope. We also identify an 8.5-day transiting planet candidate orbiting EPIC248435395, a co-moving companion to K2-266.

\end{abstract}


\section{Introduction} 
Our understanding of exoplanet demographics has rapidly expanded as a direct result of the success of the \Kepler\ and \Ktwo\ missions. With the successful launch of the Transiting Exoplanet Survey Satellite ({\it TESS}) mission, which is expected to discover thousands of new exoplanetary systems \citep{Ricker:2015}, the community is now focused on understanding the mechanisms responsible for the diversity of exoplanet architectures. We now know of over 700 multi-planet systems and a total of more than 3700 confirmed or validated planets to date\footnote{\url{https://exoplanetarchive.ipac.caltech.edu/}}. From these discoveries, we know that the most commonly known planets with periods P$<$100 days are smaller than Neptune, a large fraction of which are super-Earths and mini-Neptunes \citep[R$_P$ = 1.5 -- 4 \re;][]{Fressin:2013}. With no analogues in our own Solar System, our understanding of these planets is limited. 

\begin{figure*}[!ht]
\vspace{0.3in}
\centering\includegraphics[width=0.95\linewidth, trim = 0 5.8in 0 0]{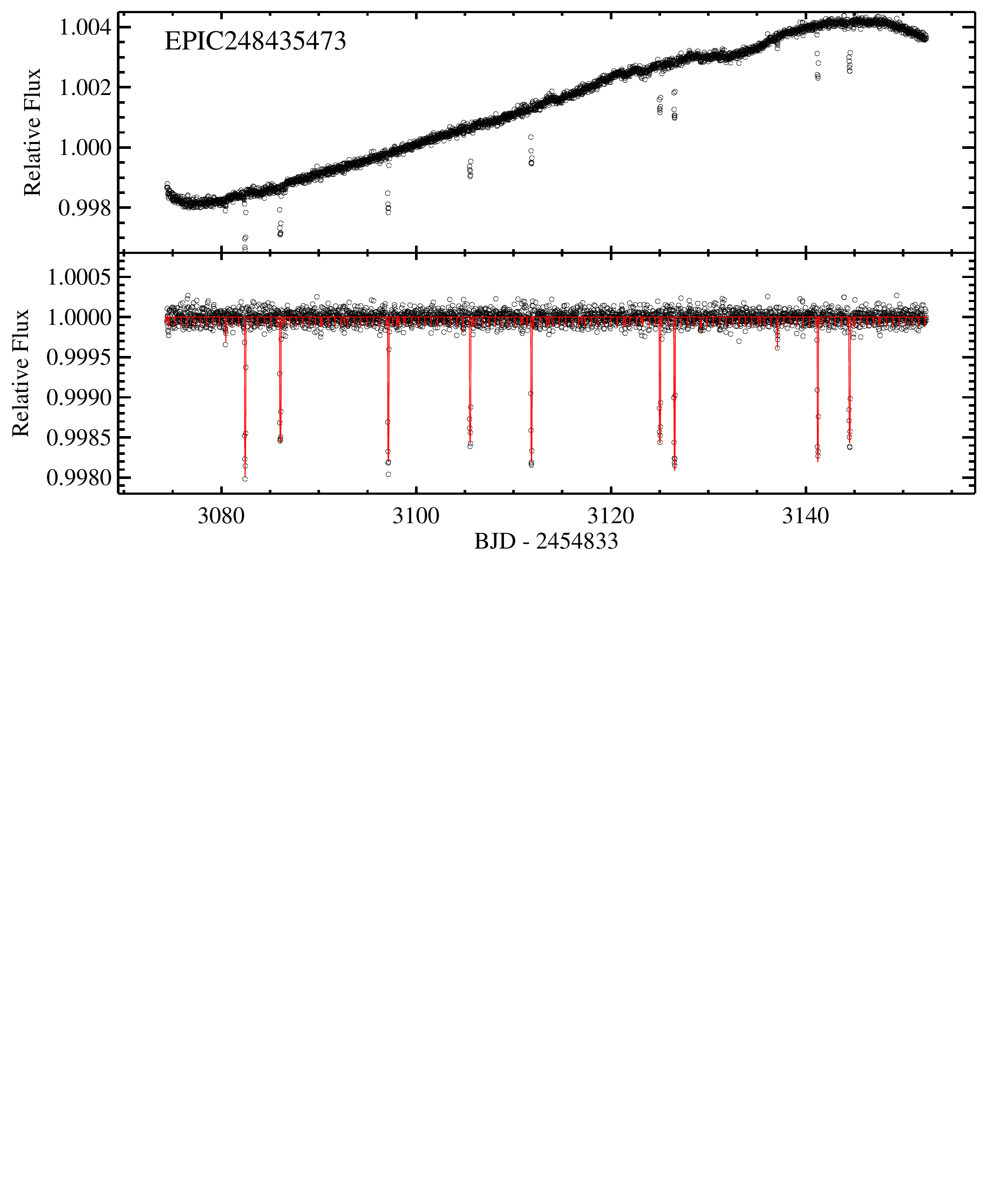}
\caption{(Top) The full K2 light curve of \thisstar from Campaign 14, corrected for systematics using the technique described in \citet{Vanderburg:2014} and \citet{Vanderburg:2016b}. The observations are plotted in open black circles, and the best fit models are plotted in red. (Bottom) The flattened final {\it K2} light curve  used in the EXOFASTv2 fit.}
\label{figure:LC}
\end{figure*}

The large number of multi-planet systems discovered may provide key information to facilitate our understanding of the formation of our own Solar System. From studying multi-planet systems using \Kepler\ data, it has been determined that $\sim$30\% of Sun-like stars have compact and closely aligned architectures, with planetary radii R$_P$>\re\ and orbital periods less than 400 days \citep{Zhu:2018}. Planets in systems with {\it large} mutual inclinations, however, might not all transit. The mutual inclination could be dependent on the number of planets in the system. Specifically, systems with fewer planets have larger mutual inclinations, possibly explaining the observed excess of Kepler single planet systems \citep{Zhu:2018}.  If unaccounted for, this bias can affect our understanding of planet formation. Fortunately, typical mutual inclinations within transiting systems can be constrained by measuring the ratio of transit durations of adjacent transiting planets. Studies that constrain the underlying multiplicity and distribution of inclinations suggest that transiting multi-planet systems are close to aligned, with mutual inclinations typically less than a few degrees \citep{Fang:2012, Figueira:2012, Swift:2013, Fabrycky:2014, Ballard:2016}. However, many studies have shown that the observed population is not well represented by a single-component model \citep{Lissauer:2011, Ballard:2016}, and this claim is supported by simulations of late-stage planet formation \citep{hansen:2013}; the underlying population may consist of some well-aligned systems and some with large mutual inclinations. 

Ultra Short-period Planets (USPs), planets that orbit with periods less than a day, may provide insight into the origin of mutually misaligned planetary systems. These are relatively rare objects (0.5\% of all stars, \citealp{Sanchis-Ojeda:2014}), but their close proximity to their host star allows them to transit at a larger range of inclinations relative to our line of sight. This relatively high transit probability makes the USP in a multi-planet system more likely to transit than the longer-period planets in the system (e.g., 55 Cancri, \citealp{Fischer:2008, Batalha:2011}). It also makes it more likely that \textit{misaligned} systems containing USPs will be observed to host multiple transiting planets, which affords greater opportunities for detailed investigations of the physical and dynamical properties of the planets. USPs therefore have the potential to help us understand the origin of planetary systems with high mutual inclinations. 

Since young stars are larger in radius than their sizes on the main sequence, by factors of 3 -- 4, it is unlikely that USPs could form in situ: the host star would have engulfed some of the known USPs based on stellar properties derived from pre-main-sequence evolutionary tracks \citep{Palla:1991, D'Antona:1994}. As a result, one possible origin scenario is that USP  migration is influenced by gravitational interactions with other planets or stars, increasing the planet's orbital eccentricity. This "High Eccentrictiy Migration" mechanism (HEM), has also been proposed to explain the origin of hot Jupiters \citep[see, e.g.,]{Petrovich:2018}. These systems initially retain their primordial eccentricities from these interactions \citep{Rasio:1996, Wu:2003, Fabrycky:2007,Nagasawa:2011, Wu:2011}, but subsequent tidal interactions should circularize the orbit (e.g., \citealt{Adams:2006}). However, the inclination excited by HEM may remain, resulting in highly misaligned planetary orbits.  

Another possible explanation for misaligned planetary systems is that they originate from misaligned disks around young stars. It is known that young stars are surrounded by circumstellar disks of gas and dust that eventually form the planetary systems that are observed today. From high resolution observations of these circumstellar disks, for example using the Atacama Large Millimeter/submillimeter Array (ALMA), we know that these disks are not smooth and uniform. Instead they contain gaps or rings \citep{ALMA:2015}, and display misalignment with their disks and even multiple disks (e.g., see Beta Pic, \citealp{Heap:2000}). Additionally, wide binary systems where each star has its own circumstellar disk have been shown to be mutually misaligned \citep[e.g.,][and references therein]{Jensen:2014}.

\begin{figure*}[!ht]
\vspace{0.3in}
\centering\includegraphics[width=0.64\linewidth, trim = 0.0 5.9in 0.0 0in]{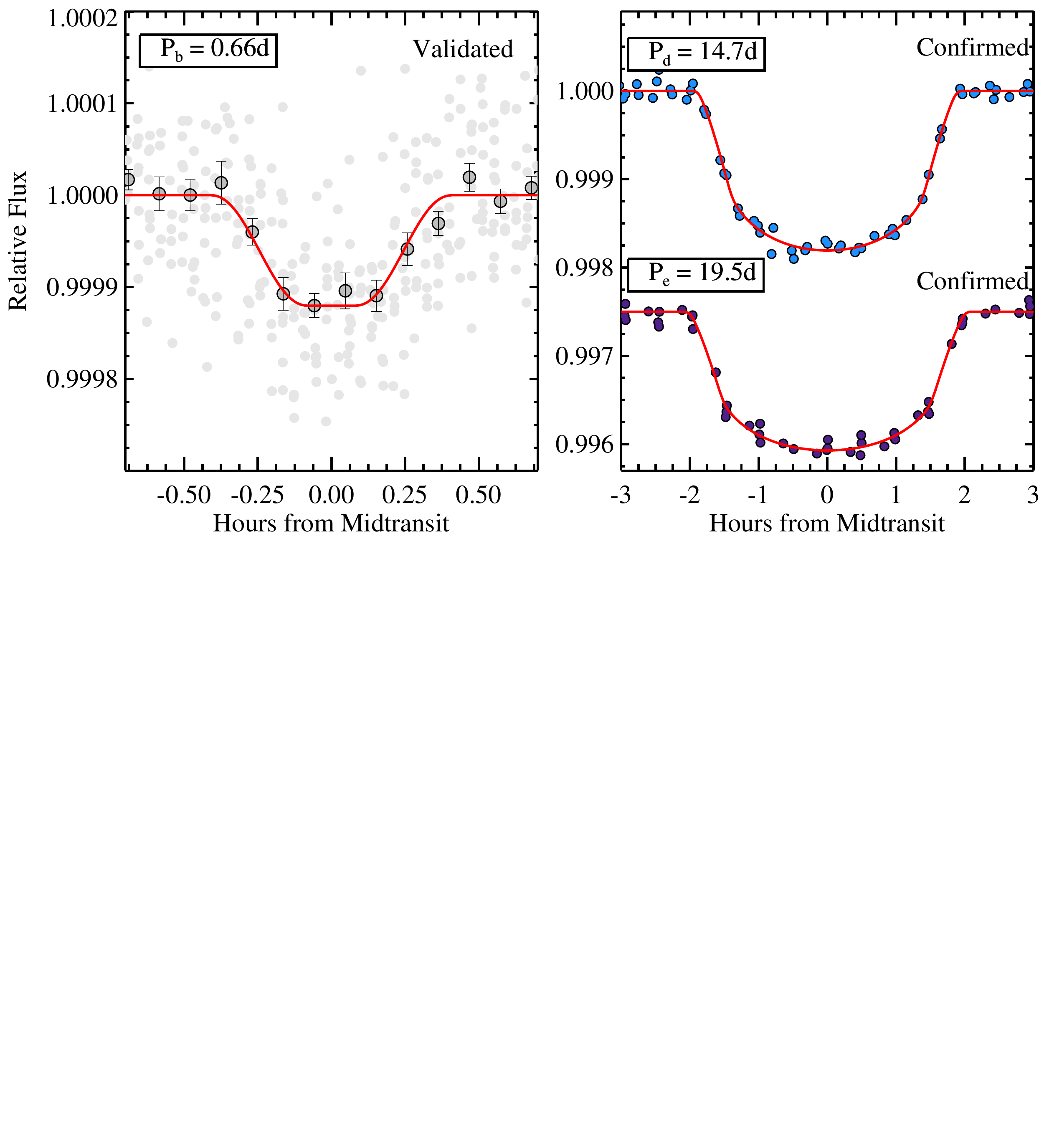}\includegraphics[width=0.36\linewidth, trim = 0.0 2.1in 0.0 1in]{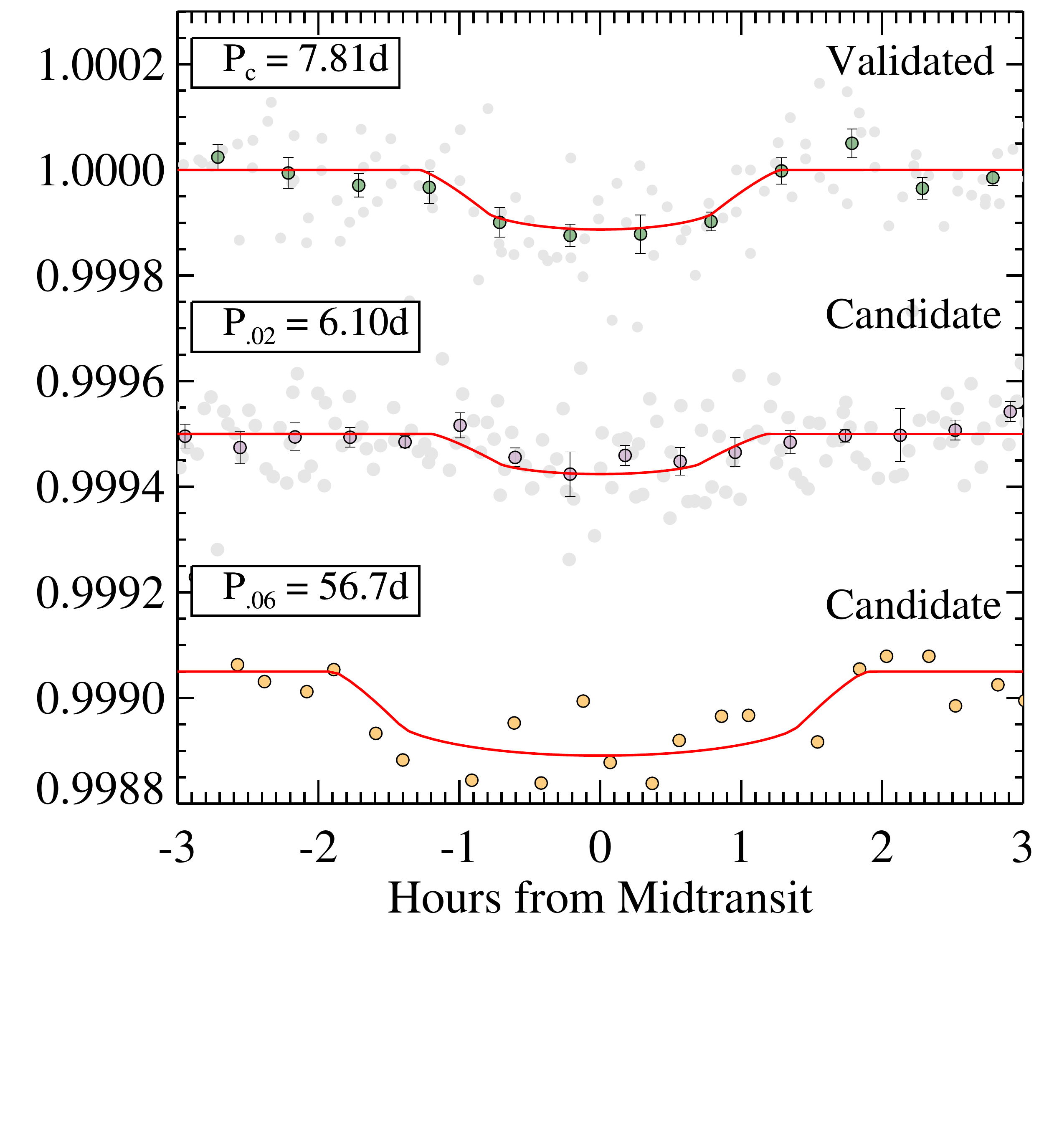}
\caption{ The phase-folded corrected {\it K2} light curve  for the four validated planets in the system \thisstar b (Left), d (middle), e (middle), and c (right), and the two additional planet candidates .02 (right) and .06 (right). For planets b, d, and candidate c, the full phasefolded LC is shown in light gray and the binned points are shown in color with error bars. The red line corresponds to the final EXOFASTv2 transit model. }
\label{figure:LCplanets}
\end{figure*}

Using high precision photometric observations from {\it Spitzer} and \Ktwo\, there has been a sub-class of young stellar objects identified called ``dippers'' that display large amplitude ($>$10\%) dimming events that occur on timescales of days \citep{Alencar:2010, Morales:2011, Cody:2014, Ansdell:2016A}. The observed variability has been attributed to extinction by dust in the inner disk, implying that disks would need to be relatively edge-on, as suggested for the archetypal dipper AA Tau \citep{Bouvier:1999}. However, recent high spatial resolution millimeter mapping of AA Tau by ALMA shows a modestly inclined disk at $59.1^{\circ}$ \citep{Loomis:2017}. Even more extreme examples exist, such as the dipper J1604-2130, for which ALMA observations reveal the disk to be nearly face-on \citep{Ansdell:2016B}. These observations, combined with the photometric dimming events observed suggest that the inner disk is more aligned to our line of sight, and therefore, misaligned relative to the outer disk. Finally, we note that molecular cloud cores that form stars do not have perfectly well-ordered distributions of angular momentum, so that the formation of disks, and later planets, naturally produces some mis-alignment (e.g., see \citealt{spalding2014} and references therein). 

Multi-planet systems also allow us to determine key physical planet parameters, such as mass and orbital eccentricity, through the detection and analysis of transit timing and duration variations (TTV \& TDV, respectively, see  \citealp{Agol:2005,Holman:2005}). The slight variations in the timing between consecutive transits are caused by another planet in the system, and result from exchanging energy and angular momentum due to their mutual gravitational interaction. Systems that have planets in or near mean motion resonance (MMR) can produce large amplitude timing variations, allowing the measurement of mass and eccentricity for small planets with longer periods. Efforts to analyze the TTVs for a large sample of planetary systems have provided mass and eccentricity measurements for planets that would not be accessible from other techniques, such as radial velocities \citep{Steffen:2013, Holczer:2016, Jontof:2016, Hadden:2017}.  

In this paper, we present the discovery and characterization of a compact multi-planet system orbiting the late K-star \thisstar. Using observations from the \Ktwo\ mission, we have identified up to six planets orbiting \thisstar, with periods of 0.66, 6.1, 7.8, 14.7, 19.5, and 56.7 days. We are able to confidently confirm the planetary nature of two of these planets (P$_d$ = 14.7 days \& P$_e$ = 19.5 days), validate two more as planets (P$_b$ = 0.66 days \& P$_c$ = 7.8 days), and we classify the other two (weaker) signals as planetary candidates. From a simultaneous global model of all six planets and candidates, we find that the orbit of \thisstar b has an inclination of 75.32\degr, while the other five planets and candidates have inclinations of 87\degr\ to 90\degr. This significant misalignment of the inner planet has interesting implications for the dynamical history of the system, and may suggest that it had a different evolutionary path than the rest of the planets. Additionally, \thisstar has a co-moving companion, \thisstarcomp, that is 42$\arcsec$ away and an early M-star. This companion was resolved by \Ktwo, and we report the identification of a planet candidate orbiting \thisstarcomp with a period of 8.5d. 


The paper is organized in the following way: We first discuss our photometric and spectroscopic observations in \S\ref{Obs}. Our EXOFASTv2 global model methodology and results are then presented in \S\ref{sec:GlobalModel}. We present all observations on the co-moving companion (\thisstarcomp) and discuss the nature of the companion star and its planetary candidate in \S\ref{sec:companion}. A dynamical analysis of the system is carried out in \S\ref{sec:dynamics}. Finally, we discuss our results in \S\ref{sec:discussion} and conclusions in \S\ref{sec:conclusion}. 

\begin{deluxetable*}{llccc}
\tablecaption{\thisstar\ \& \thisstarcomp Magnitudes and Kinematics}
\startdata
\hline
Other identifiers\dotfill & & & \\
       
	  & &{\thisstar}& \\
	  & &{2MASS J10314450+0056152}&{2MASS 10314174+0056048} \\
	  & &{EPIC248435473}&{\thisstarcomp}\\
\hline
\hline
Parameter & Description & Value & Value & Source\\
\hline 
$\alpha_{J2000}$\dotfill	&Right Ascension (RA)\dotfill & 10:31:44.506&10:31:41.749& 1	\\
$\delta_{J2000}$\dotfill	&Declination (Dec)\dotfill & +00:56:15.27&+00:56:04.94& 1	\\
\\
$B$\tablenote{The uncertainties of the photometry have a systematic error floor applied. Even still, the global fit requires a significant scaling of the uncertainties quoted here to be consistent with our model, suggesting they are still significantly underestimated for one or more of the broad band magnitudes}\dotfill		& APASS Johnson $B$ mag.\dotfill	& 13.001 $\pm$	0.02& 15.011 $\pm$ 0.04		& 2	\\
$V$\dotfill		& APASS Johnson $V$ mag.\dotfill	& 11.808 $\pm$	0.02& 13.538 $\pm$ 0.02		& 2	\\
$G$\dotfill     & Gaia $G$ mag.\dotfill     &11.3527$\pm$0.0009&12.7285$\pm$0.0005& 7,8\\
$g'$\dotfill		& APASS Sloan $g'$ mag.\dotfill	& 12.407 $\pm$ 0.02	& 14.270$\pm$0.03	& 2	\\
$r'$\dotfill		& APASS Sloan $r'$ mag.\dotfill	& 11.311 $\pm$ 0.02	& 12.931$\pm$0.03	& 2	\\
$i'$\dotfill		& APASS Sloan $i'$ mag.\dotfill	& 10.927 $\pm$ 0.04 & 12.111$\pm$0.03	 & 2	\\
\\
J\dotfill			& 2MASS $J$ mag.\dotfill & 9.611  $\pm$ 0.05	& 10.405$\pm$0.03	& 3, 4	\\
H\dotfill			& 2MASS $H$ mag.\dotfill & 9.041 $\pm$ 0.03	    &  9.784$\pm$0.02	& 3, 4	\\
K$_S$\dotfill			& 2MASS $K_S$ mag.\dotfill & 8.897 $\pm$ 0.02& 9.562$\pm$0.02	& 3, 4	\\
\\
\textit{WISE1}\dotfill		& \textit{WISE1} mag.\dotfill & 8.805 $\pm$ 0.022 & 9.439 $\pm$ 0.022		& 5	\\
\textit{WISE2}\dotfill		& \textit{WISE2} mag.\dotfill & 8.897 $\pm$ 0.02 & 9.416 $\pm$ 0.019		& 5 \\
\textit{WISE3}\dotfill		& \textit{WISE3} mag.\dotfill &  8.787 $\pm$ 0.02 & 9.316 $\pm$ 0.034		& 5	\\
\textit{WISE4}\dotfill		& \textit{WISE4} mag.\dotfill & 8.789 $\pm$0.437 & 8.996 $\pm$ 0.508		& 5	\\
\\
$\mu_{\alpha}$\dotfill		& Gaia DR2 proper motion\dotfill & 56.871 $\pm$ 0.151 & 53.231 $\pm$ 0.161		& 7, 8 \\
                    & \hspace{3pt} in RA (mas yr$^{-1}$)	& & \\
$\mu_{\delta}$\dotfill		& Gaia DR2 proper motion\dotfill 	&  -68.828 $\pm$ 0.242 & -72.735 $\pm$ 0.263     &  7,8 \\
                    & \hspace{3pt} in DEC (mas yr$^{-1}$) & & \\
$\pi$\dotfill & Gaia Parallax (mas) \dotfill & 12.87 $\pm$ 0.06 & 12.85 $\pm$ 0.06 & 7,8 \\
$RV$\dotfill & Systemic radial \hspace{9pt}\dotfill  & $10.848 \pm 0.066$ &    $12.84 \pm 0.31$ &\S\ref{sec:TRES},\S\ref{sec:companion} \\
     & \hspace{3pt} velocity (\kms)  & & \\
\enddata
\tablecomments{
References are: $^1$\citet{Cutri:2003},$^2$\citet{Henden:2016},$^3$\citet{Cutri:2003}, $^4$\citet{Skrutskie:2006}, $^5$\citet{Cutri:2014}, $^6$\citet{Zacharias:2017},$^7$\citet{Gaia:2016}, $^8$\citet{Gaia:2018}
}

\label{tab:LitProps}
\end{deluxetable*}

\section{Observations, Archival Data, and Validation}
\label{Obs}

\subsection{K2 Photometry}
Since the failure of the second reaction wheel, the \Kepler\ spacecraft has been re-purposed to observe a set of fields along the ecliptic. Each \Ktwo\ campaign lasts $\sim$80 days \citep{Howell:2014}, achieving similar precision to the original \Kepler\ mission \citep{Vanderburg:2016b}. \thisstar was observed during \Ktwo\ Campaign 14 from UT 2017 Jun 02 until UT 2017 Aug 19, obtaining 3504 observations on a 30 minute cadence (see Figure \ref{figure:LC}). Following the strategy described in \citet{Vanderburg:2014} and \citet{Vanderburg:2016b}, the light curves were extracted from the \Kepler-pipeline calibrated target pixel files from the Mikulski Archive for Space Telescopes\footnote{MAST; \url{https://archive.stsci.edu/}}, corrected for the \Ktwo\ spacecraft-motion-induced systematics, and searched for transiting planet candidates. From our search of \thisstar, we identified three super-Earth/sub-Neptune sized transiting exoplanet candidates with periods of 0.66, 14.7, and 19.5 days with signal-to-noise (S/N) values of 13.0, 114.6, and 111.5. In addition, some of us (MHK, MO, HMS, IT) performed a visual inspection of the light curve using the LCTOOLS\footnote{\url{https://sites.google.com/a/lctools.net/lctools/home}} software \citep{Kipping:2015}. From this visual inspection, we identified two additional Earth sized exoplanet candidates with periods of 6.1 and 7.8 days with S/N values of 8.3 and 10.6. An additional visual inspection of the K2 light curve  led to the identification of a sixth planet candidate at 56.7 days with a S/N value 6.6. The phase-folded light curves for each planet candidate is shown in Figure \ref{figure:LCplanets}. We note that the two transits of this candidate overlap with other candidates in the system. The K2 light curve was reprocessed where all six planets were simultaneously fit along with the stellar variability and known \Ktwo\ systematics. The corresponding light curve  was flattened by dividing out the best-fit stellar variability using a spline fit with breakpoints every 0.75 days. The final light curve  for \thisstar, shown in Figure \ref{figure:LC}, has a 30 minute cadence noise level of 70 ppm, and a 6 hour photometric precision of 19 ppm.


\subsection{TRES Spectroscopy} \label{sec:TRES}

Using the Tillinghast Reflector Echelle Spectrograph \citep[TRES;][]{furesz:2008}\footnote{\url{http://www.sao.arizona.edu/html/FLWO/60/TRES/GABORthesis.pdf}} on the 1.5 m Tillinghast Reflector at the Fred L. Whipple Observatory (FLWO) on Mt. Hopkins, AZ we obtained 8 observations of \thisstar\ between UT 2017 Nov 23 and UT 2018 Apr 10. TRES has a resolving power of $\lambda / \Delta \lambda = 44000$, and an instrumental radial velocity (RV) stability of $10$--$15$ \ms. The spectra were optimally extracted, wavelength calibrated, and cross-correlated to derived relative RVs following the techniques described in \citep{Buchhave:2010}. We cross-correlate each spectrum, order by order, against the strongest observed spectrum, and fit the peak of the cross-correlation function summed across all orders to derive the relative RVs. Uncertainties are determined from the scatter between orders for each spectrum. We use RV standard stars to track the instrumental zero point over time, and apply these zero point shifts (typically $<15$\ \ms) to the relative RVs and propagate uncertainties in the zero point shifts to the RVs. This is why the strongest spectrum, correlated against itself, does not have an RV of 0 \ms. The final relative RVs are given in Table \ref{tab:rv}. Using the RV standards to set the absolute zero point of the TRES system, we also determine the RV of \thisstar on the IAU standard system to be 10.848 $\pm$ 0.066 \kms, where the uncertainty is dominated by the uncertainty in the shift from relative to absolute RV.

\begin{deluxetable}{l l l l}[bt]
\tabletypesize{\scriptsize}
\tablecaption{Relative Radial Velocities for \thisstar and \thisstarcomp \label{tab:rv}}
\tablewidth{0pt}
\tablehead{
\colhead{\bjdtdb} & \colhead{RV (m s$^{-1}$)} & \colhead{$\sigma_{RV}$ (m s$^{-1}$)} & \colhead{Target}
}
\startdata
2458081.028322 &	-30.1 &	17.7 & \thisstar \\
2458090.980928 &	 -2.5 &	32.0 & \thisstar \\
2458106.965261 &	 -3.2 &	43.6 &	\thisstar \\
2458107.923621 &	-55.5 &	28.2 & \thisstar \\
2458211.644609 &	-26.3 &	21.2 &	\thisstar \\
2458212.644636 &	-36.5 &	25.6 &	\thisstar \\
2458213.663584 &	-23.9 &	33.3 &	\thisstar \\
2458218.808648 &	-30.0 &	32.5 &	\thisstar \\
\hline
2458156.911319 &	425.9 &	46.9 & \thisstarcomp\\
2458211.675712 &	-21.6 &	47.9 & \thisstarcomp\\
\enddata
\end{deluxetable}

\subsection{Palomar TripleSpec Observations}
\label{sec:TSPEC}
We refined the characterization of \thisstar by acquiring near-infrared spectra using TripleSpec on the 200" Palomar Hale telescope on 1 December 2017. TripleSpec has a fixed slit of 1" x 30" slit, enabling simultaneous observations across J, H, and K bands (1.0 - 2.4 microns) at a spectral resolution of 2500-2700 \citep{Herter:2008}. Following \citet{Muirhead:2014}, we obtained our observations using a 4-position ABCD not pattern to reduce the influence of bad pixels on our resulting spectra. As in \citet{Dressing:2017}, we reduced our data using a version of the publicly available {\tt Spextool} pipeline (Cushing et al. 2004) that was modified for use with TripleSpec data (available upon request from M. Cushing). We removed telluric contamination by observing an A0V star at a similar airmass and processing both our observations for both the A0V star and \thisstar with the {\tt xtellcor} telluric correction package \citep{Vacca:2003}. 

After reducing the spectra, we estimated stellar properties by applying empirical relations developed by \citet{Newton:2014, Newton:2015} and \citet{Mann:2013a, Mann:2013b}. Specifically, we estimated the stellar effective temperature and radius by measuring the widths of Al and Mg features using the publicly available, IDL-based {\tt tellrv} and {\tt nirew} packages \citep{Newton:2014, Newton:2015}. We then employed the stellar effective temperature-mass relation developed by \citet{Mann:2013b} to infer the stellar mass from the resulting stellar effective temperature estimate. We also estimated the stellar metallicity ([M/H] and [Fe/H]) using the relations developed by \citet{Mann:2013a}. For more details about our TripleSpec analysis methods, see \citet{Dressing:2017}, Dressing et al., (in prep). 

The resulting stellar properties were $T_{\rm eff} = 4192 \pm 77$ K, $M_\star = 0.67^{+0.08}_{-0.07} \msun$, and $R_\star = 0.63 \pm 0.03$. This values for the stellar mass is consistent with those estimated from our EXOFASTv2 analysis (see Table 3). However, the radius is $\sim$3$\sigma$ different from the EXOFASTv2 fit using the broadband photometry and Gaia DR2 parallax.

\subsection{Archival ``Patient'' Imaging}
\label{sec:ArchivalImaging}
To check for nearby stars (either physically associated companions or coincidental alignments) that may influence our results, we examined archival observations from National Geographic Society Palomar Observatory Sky Survey (NGS POSS) from 1952. The proper motion of \thisstar is $\mu_{\alpha}$ = 56.9 mas and $\mu_{\delta}$ = -68.8 mas, and has moved $\sim$6$\arcsec$ in the 66 years since the original POSS observations were taken. The present-day position of \thisstar\ is located right at the edge of the saturated point-spread-function of \thisstar\ in the original POSS plates. While the present-day position of \thisstar\ is not completely resolved in the POSS image, if there was a bright-enough background star at the present-day position of \thisstar, we would expect to see some elongation of the POSS point spread function at that position. We see no evidence for such an elongation in POSS plates with either a red-sensitive or blue-sensitive emulsion. We estimate that we can rule out background stars at the present-day position of \thisstar\ down to a magnitude of about 19 in blue, and a magnitude of about 18 in the red. Figure \ref{figure:AO} shows our archival imaging overlaid with the K2 photometric aperture used to extract the light curves. 

We used modern imaging from the Pan-STARRS data release to search for faint companions at distances greater than a few arcseconds from \thisstar\ \citep{Flewelling:2016}. In the Pan-STARRS images, we identified one star located inside our best photometric aperture about 9 magnitudes fainter than \thisstar. In principle, if this star were a fully eclipsing binary (with 100\% deep eclipses\footnote{While the greatest eclipse depth possible from two main-sequence eclipsing binaries is about 50\% (caused by an equal-brightness binary), we also consider the worst-case scenario of 100\% deep eclipses caused by, for example, a bright, hot white dwarf being eclipsed by a cool M-dwarf or brown dwarf \citep[e.g.][]{rappaport2018}.}), it could contribute a transit-like signal to the light curve of \thisstar\ with a depth of at most about 250 ppm. This is shallower than the transits of the two sub-Neptunes, but could in principle contribute the transits of the other four candidates. We therefore extracted the K2 light curve from a smaller aperture (shown in Figure \ref{figure:AO} as a navy blue outline overlaid on the Pan-STARRS image of \thisstar), which excludes the companion star detected in Pan-STARRS imaging. We find in the noisier light curve extracted from the smaller aperture, the transits of the ultra-short-period and 7.8d planets are convincingly detected, but the transits of the two weaker candidates (at 6.1 and 56.7 days) do not convincingly appear (due to the increased noise in the light curve). We therefore cannot rule out a blended background eclipsing binary origin for at least one of those candidates.

\subsection{Keck/NIRC2 AO Imaging}
\label{sec:AO}
\begin{figure*}[!ht]
\centering\includegraphics[width =0.9\linewidth,angle =0]{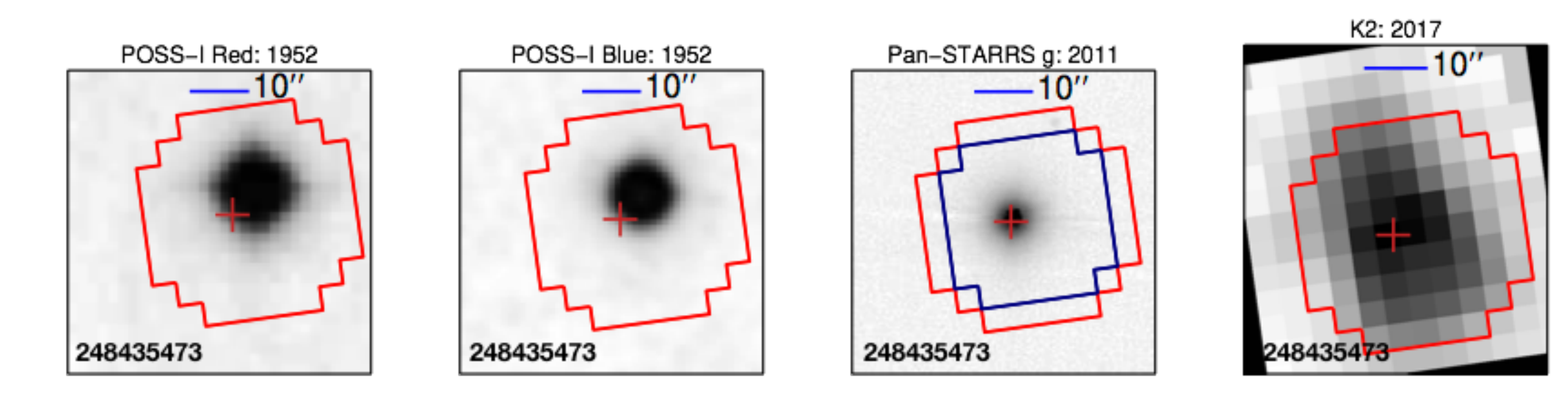}
\vspace{-0.25in}

\centering\includegraphics[width =0.35\linewidth,angle =90, trim = 0in 0in 0in 0.75in]{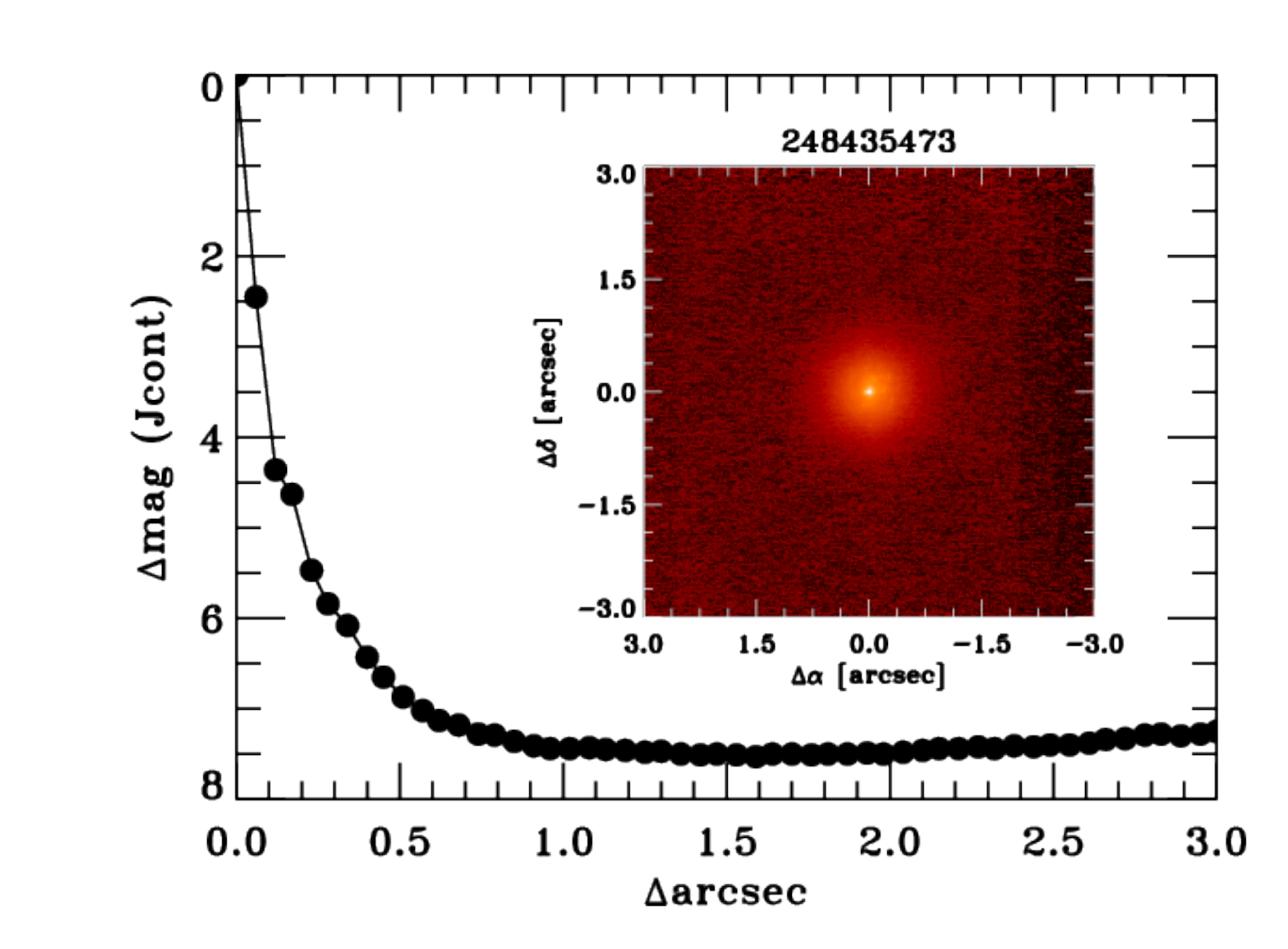}\centering\includegraphics[width=0.35\linewidth, angle =90, trim = 0 0in 0 0.75in]{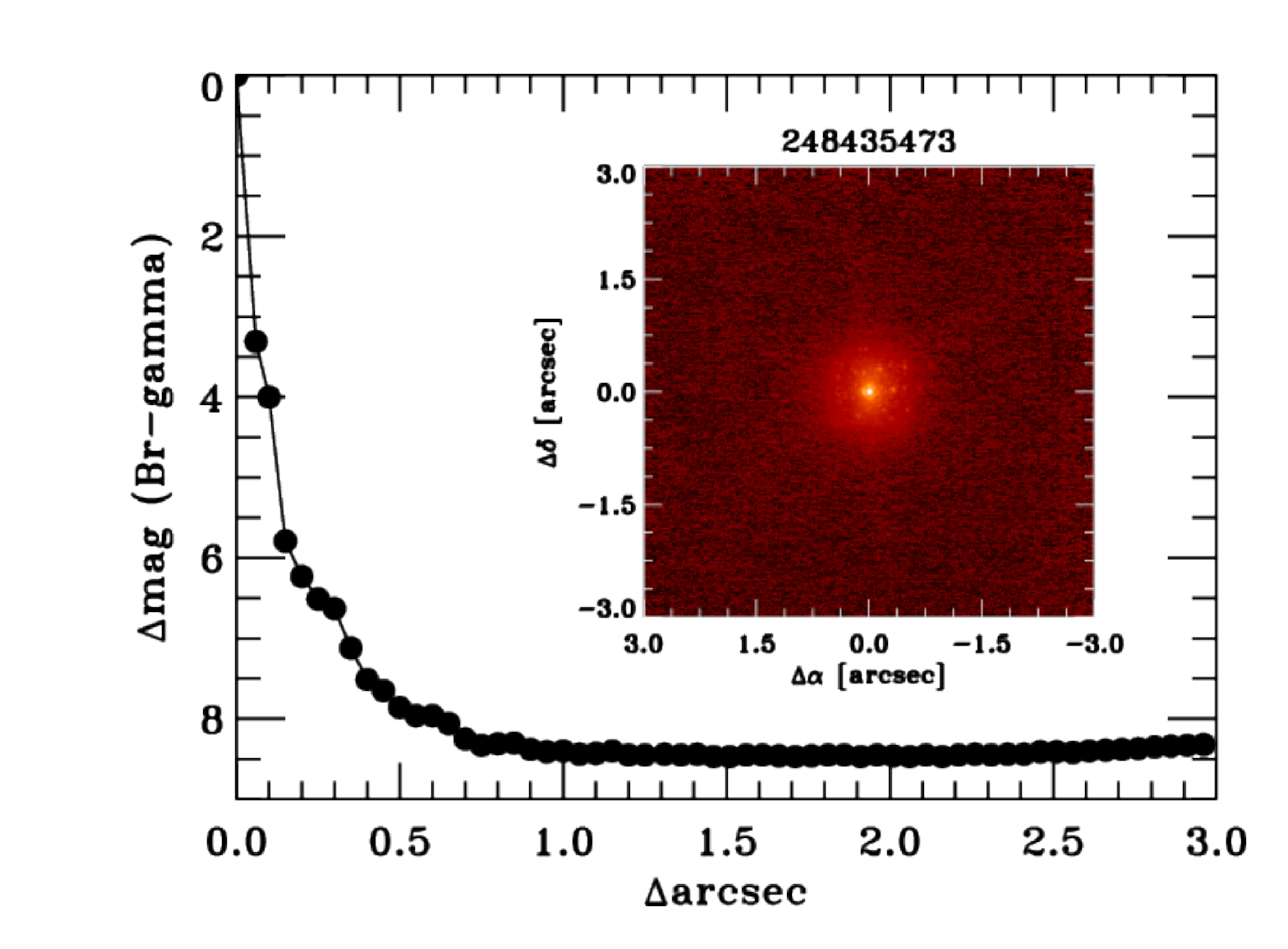}
\vspace{-0.25in}

\centering\includegraphics[width =0.9\linewidth,angle =0]{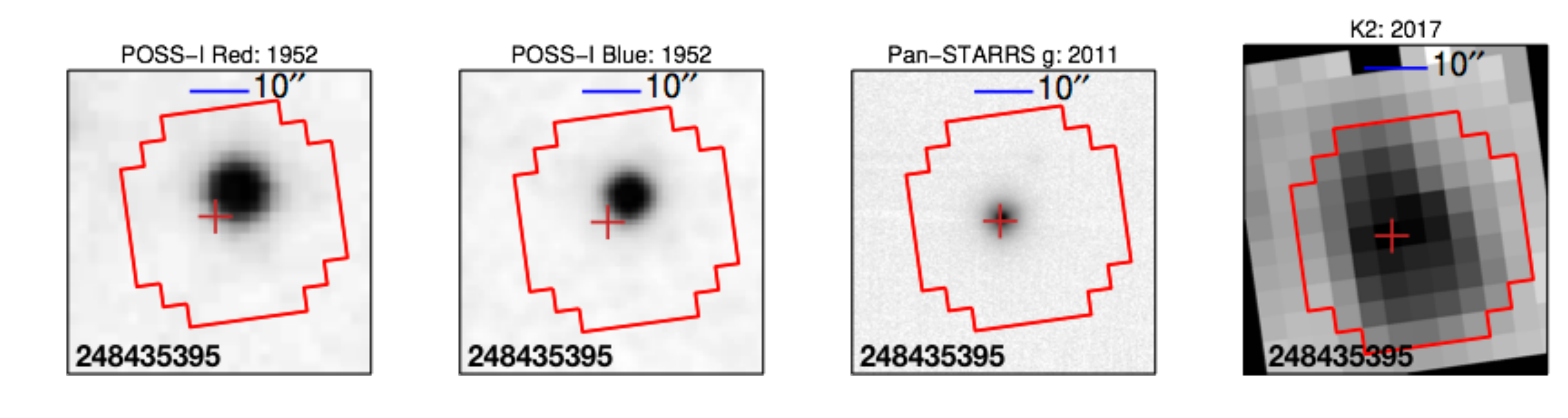}
\vspace{-0.25in}

\centering\includegraphics[width =0.35\linewidth,angle =90, trim = 0in 0in 0 0.75in]{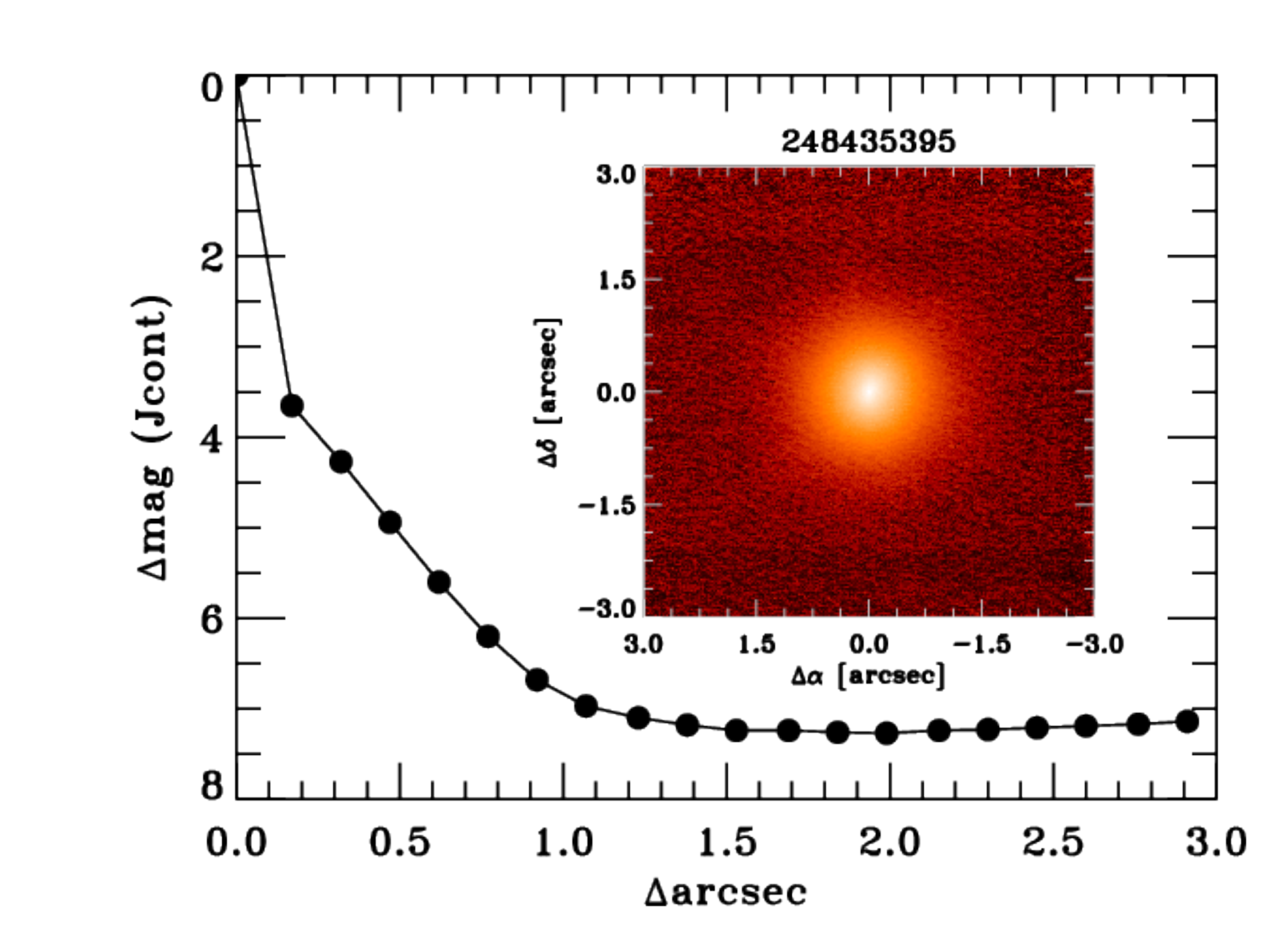}\centering\includegraphics[width=0.35\linewidth, angle =90, trim = 0 0in 0 0.75in]{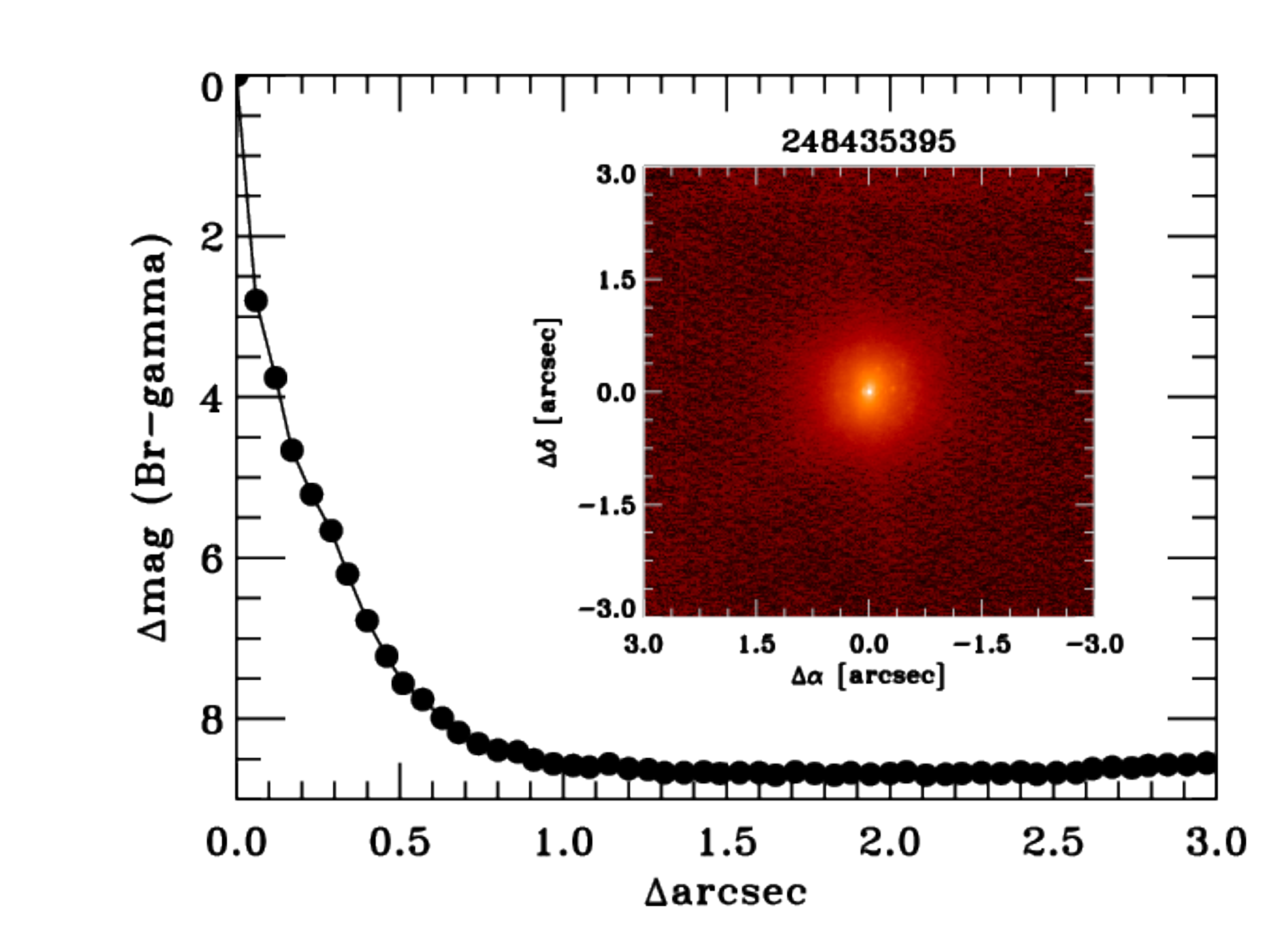}
\caption{Archival imaging from the National Geographic Society Palomar Observatory Sky Survey (NGS POSS) of \thisstar\ (1st Row) taken with a (1st panel) red and (2nd panel) blue emulsion in 1952. (3rd panel) Archival imaging from the Pan-STARRS survey of \thisstar\ taken in 2011. (Top Right) Summed image of \thisstar\ from K2 observations. The aperture selection is described in \citet{Vanderburg:2016b}. (Bottom Row) The same four images for \thisstarcomp. The Keck (left) $J$-band and (right) Br-$\gamma$ contrast curve and image (inset) of (Second Row) \thisstar and (Third Row) \thisstarcomp.  We note that the $J$-band image for \thisstarcomp was observed through poorer atmospheric conditions leading to reduced image quality. We find no evidence of any additional components in either system. }
\label{figure:AO}
\end{figure*}

We obtained high resolution images of \thisstar using the Near Infrared Camera 2 (NIRC2) on the W. M. Keck Observatory. Two observations of each target were taken on UT 2017 December 28, one in the Br-$\gamma$ filter and the other in the $J$-band (see Figure \ref{figure:AO}). NIRC2 has 9.942 mas pix$^{-1}$ pixel scale and 1024$\times$1024 pixel array. The lower left quadrant of the array suffers from higher noise levels. To exclude this part of the detector, a 3-point dither pattern was used. The final image shown in Figure \ref{figure:AO} is created by shifting and co-adding the observations, after flat-fielding and sky subtraction. We see no other star in the 10$\arcsec$ field-of-view for \thisstar. Our sensitivity to nearby companions is determined by injecting a simulated source with a S/N of 5. The final 5$\sigma$ sensitivity curves as a function of spatial separation and the corresponding images for \thisstar in both Br-$\gamma$ and $J$ filters are shown in Figure \ref{figure:AO}.


\begin{figure*}[!ht]
\centering 
\includegraphics[width=\columnwidth]{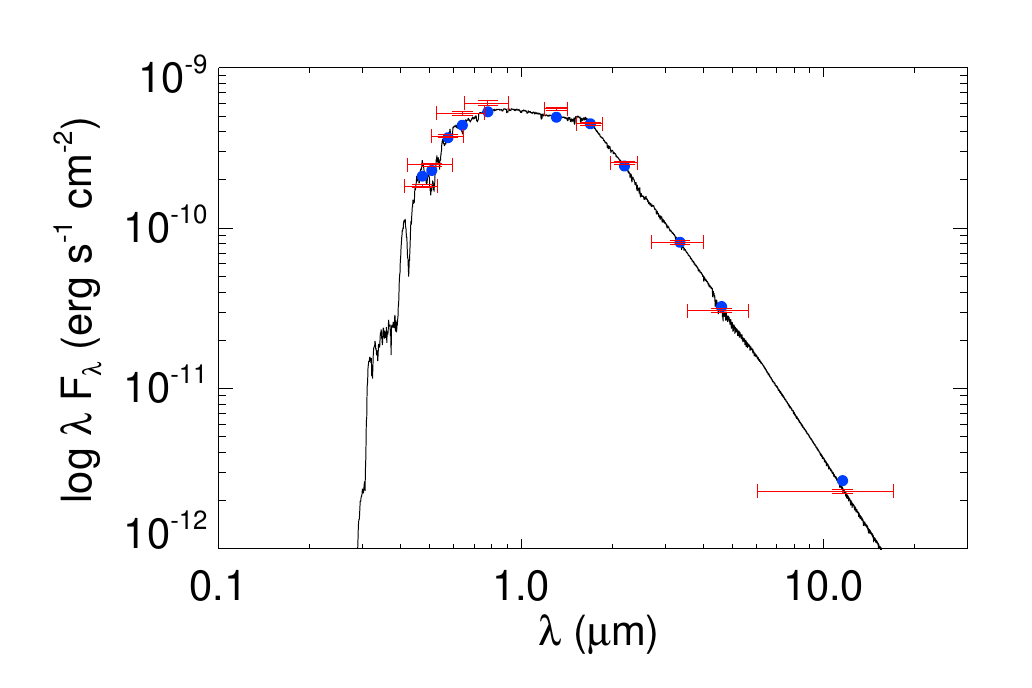}\includegraphics[width=\columnwidth]{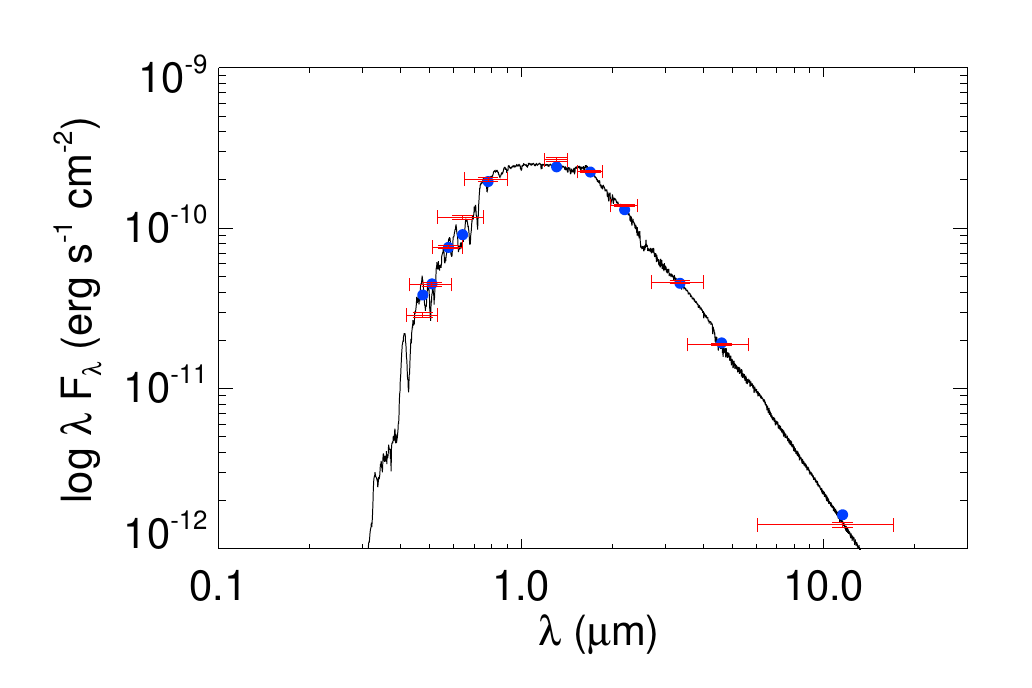}
\caption{The SED fit for (left) \thisstar and (right) \thisstarcomp from EXOFASTv2. The blue points are the predicted integrated fluxes and the red points are the observed values at the corresponding passbands. The width of the bandpasses are the horizontal red error bars and the vertical errors represent the 1$\sigma$ uncertainties. The final model fit is shown by the solid line. }
\label{fig:sed_fit}
\end{figure*}

\begin{figure}[!ht]
\centering 
\includegraphics[width=\columnwidth, , trim = 0in 0in 0 0in]{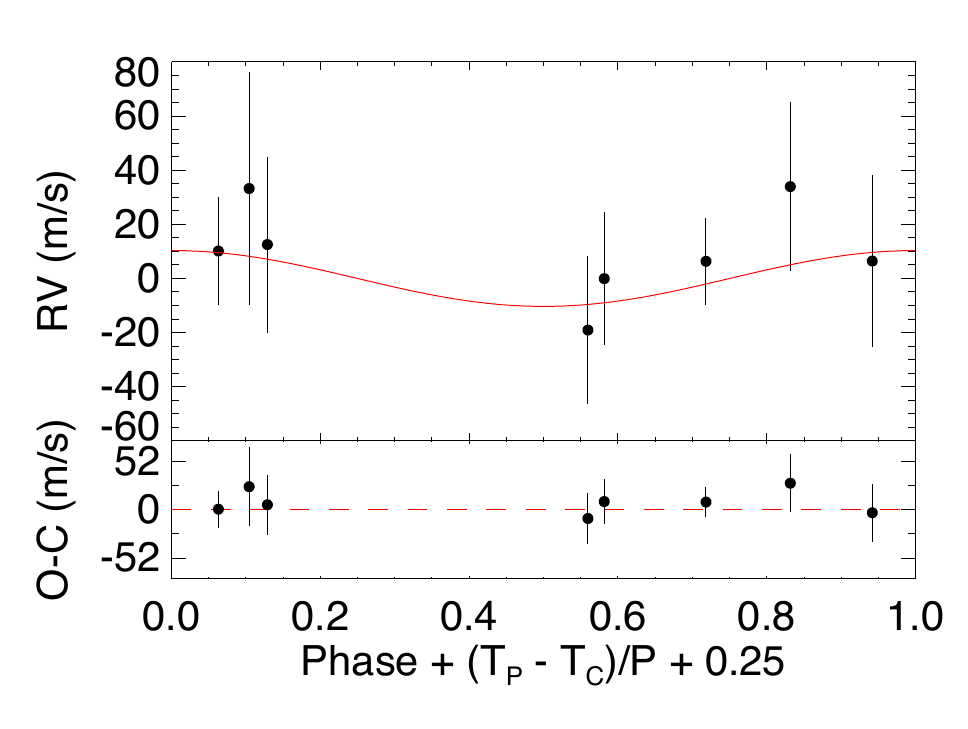}
\caption{The TRES radial velocity measurements phase-folded to the best-fit ephemeris of \thisstar b. The primary transit occurs at a phase of 0.25, where $T_P$ is the time of periastron, $T_C$ is the time of transit, and $P$ period of planet b.}
\label{fig:rv}
\end{figure}

\subsection{Statistical Validation}

We attempted to statistically validate each of the six candidates in \thisstar, a process in which the probability of planethood is estimated. If the probability is above some threshold value the candidate is upgraded to validated planet. Our method of validation followed the approach taken by \citet{Mayo:2018}. In detail, we made use of \vespa \citep{morton:2015}, a Python package based on the work of \citet{morton:2012}. \vespa calculates the false positive probability (FPP) of an exoplanet candidate by first simulating a population of synthetic stellar systems, each of which creates a transit signal due to a planet or eclipsing binary scenario. Then, \vespa calculates the FPP by determining which synthetic systems are consistent with the input observations and calculating the fraction of those systems that correspond to an eclipsing binary scenario. 

This determination is made based on inputs such as the sky position of the target, the transit signal, various stellar parameters, and contrast curves from any available high-resolution imaging. (A contrast curve relates the angular separation between the target star and an undetected companion to the maximum brightness for the putative companion.) In the case of \thisstar, we provided as input to \vespa the RA and Dec, the phase-folded light curve of the candidate in question (with transits from other candidates removed), $J$, $H$, and $K_S$ bandpass stellar magnitudes from 2MASS \citep{Cutri:2003,Skrutskie:2006} and the Kepler magnitude, stellar parameters (\teff, \logg, and \feh) calculated in Section~\ref{sec:GlobalModel}, and contrast curves from two AO images.

After we subjected each of our six candidates to validation, we made two additional adjustments to their FPP estimates. First, there are eight spectra and corresponding RV measurements collected with TRES from 2017 Nov 23 to 2018 Apr 10. The RV measurements derived from the TRES spectra did not indicate any large variations indicative of a simple eclipsing binary, so we were able to eliminate that scenario. (Note that this is different from a background eclipsing binary or hierarchical eclipsing binary scenario, which we also consider.) By eliminating the possibility of a simple eclipsing binary, the probability of the planet scenario (and each false positive scenario) was increased so that the total probability remained at unity.

Second, according to \citet{lissauer:2012}, the likelihood of one or more false positives decreases significantly when there is more than one candidate in a system. In the case of a system with more than 2 candidates, they estimate that a multiplicity boost factor of 50 is appropriate. As a result, we decreased the FPP for each candidate by a factor of 50.

After calculating FPP values for our six candidates, reducing the eclipsing binary scenario to 0 probability, and including a multiplicity boost of 50, we found final FPP values of $3.02e-05$, $7.34e-06$, $9.40e-06$, $6.80e-11$, $1.16e-12$, and $4.90e-06$ for candidates \thisstar.01, .02, .03, .04, .05, and .06. These values would each be low enough to easily validate all six candidates (e.g. \citet{Mayo:2018} used a FPP threshold value of 1e-4). However, given the inability to rule out the possibility that the faint background star we identified in Section~\ref{sec:ArchivalImaging} is an eclipsing binary, we were only able to conclusively validate candidates \thisstar.01, .03, .04, and .05, naming them \thisstar b, c, d, and e, respectively. We also refrain from validating candidates .02 and .06 because they have the lowest signal-to-noise ratios that do not pass our threshold (8.3 and 6.6, respectively). Validating such low S/N candidates is challenging because it is difficult to prove that the weakest signals detected in \Kepler\ or K2 data are astrophysical, and not the result of residual instrumental systematics or artifacts \citep{Mullally:2018}.

\section{EXOFASTv2 Global Fit for \thisstar} 
\label{sec:GlobalModel}
Using the global exoplanet fitting suite, EXOFASTv2 \citep{Eastman:2017}, we perform a simultaneous fit of the existing observations to determine the final system parameters for \thisstar. Based largely on the original EXOFAST \citep{Eastman:2013}, EXOFASTv2 provides the unique flexibility to simultaneously fit the spectral energy distribution (SED) and RV observations from multiple instruments, in combination with fitting the time series photometry for every planet in the system. Using EXOFASTv2, we simultaneously fit the flattened \Ktwo\ light curve  (accounting for the 30 minute cadence smearing, see Figure \ref{figure:LC} and \ref{figure:LCplanets}), the SED (see Table \ref{tab:LitProps}), and the radial velocity observations from TRES (see Figure \ref{fig:rv}). To characterize the host star radius within the fit, we include the the broad band photometry and Gaia DR2 parallax (See Table \ref{tab:LitProps}) \citep{Gaia:2016, Gaia:2018}. We add 0.082 mas to the DR2 parallax, as determined by \citet{Stassun:2018} and impose a systematic error floor on the uncertainty of 0.1 mas since all systematics and uncertainties should be below this \citep{Gaia:2018}. To constrain the mass of the star, we used a Gaussian prior of $0.677 \pm 0.034 \msun$ from \citet{Mann:2015}, but with the uncertainties inflated to 5\%. In a separate global fit (not reported), we used the MIST stellar isochrones \citep{Dotter:2016, Choi:2016, Paxton:2011, Paxton:2013, Paxton:2015} instead of the Gaussian prior from \citet{Mann:2015} as the primary constraint of the stellar mass and arrived at $0.748^{+0.047}_{-0.045} \msun$, a $1.3\sigma$ difference. We favor the \citet{Mann:2015} relations due to their empirical approach and the known problems with all model isochrones at low stellar masses.

Additionally, we enforce an upper limit in the $V$-band extinction ($A_V$) from the \citet{Schlegel:1998} dust maps of 0.0548 at the position of \thisstar. The final SED fit is shown in Figure \ref{fig:sed_fit}, the phase-folded RVs from TRES to planet b's period is shown in Figure \ref{fig:rv}, and the best fit transit models are shown in Figures \ref{figure:LC} and \ref{figure:LCplanets}. Given the near resonance orbit of \thisstar d and e, which would have the largest transit timing variations (TTVs), we fit for the TTVs of these two planets while fitting a linear ephemeris for planets/candidates b, .02, c, and .06. The final determined stellar and planetary parameters from our global fit are shown in Tables \ref{tab:ep248435473_0}, \ref{tab:ep248435473_1}, and \ref{tab:ep248435473_2}.

The grazing geometry of planet b means the upper limit of the planet radius is unconstrained by the light curve. However, during the global fit, we simultaneously model the radial velocities, which provide a robust upper limit on its mass. This upper limit is translated to a radius upper limit during the global fit using EXOFASTv2's integrated \citet{Chen:2017} exoplanet mass-radius relation, which excludes Jupiter-radius solutions (and even higher inclinations). Because the radial velocities are not precise enough for a measurement, the prior, which is uniform in $log(K)$, can have a significant impact on the posteriors for the RV-semi amplitude, mass, radius, and inclination of planet b and tends to favor smaller planets and smaller inclinations (in line with our prior expectation that such planets are intrinsically more numerous).

Because the \citet{Chen:2017} relations only use a sample of planets with robustly detected masses and radii, and we can typically measure robust radii for smaller planets than we can measure the corresponding masses, there is likely a selection effect in the \citet{Chen:2017} relations that bias it toward larger masses at low signal to noise. As a consequence, for a given mass, we expect to over-estimate the radius. Our radius upper limit would likely be somewhat smaller if we used a relation that accounted for non-detections within our fit.

The \citet{Weiss:2014} exoplanet mass-radius relations are an often used alternative which attempts to account for the bias from non-detections. However, they only apply to rocky planets ($R_p < 4\re$), and so could not be used to exclude large planets, whereas the \citet{Chen:2017} relations are defined and continuous from rocky planets to stars.

\begin{figure}[!ht]
\centering 
\includegraphics[width=\columnwidth, trim = 0in 0in 0 0in]{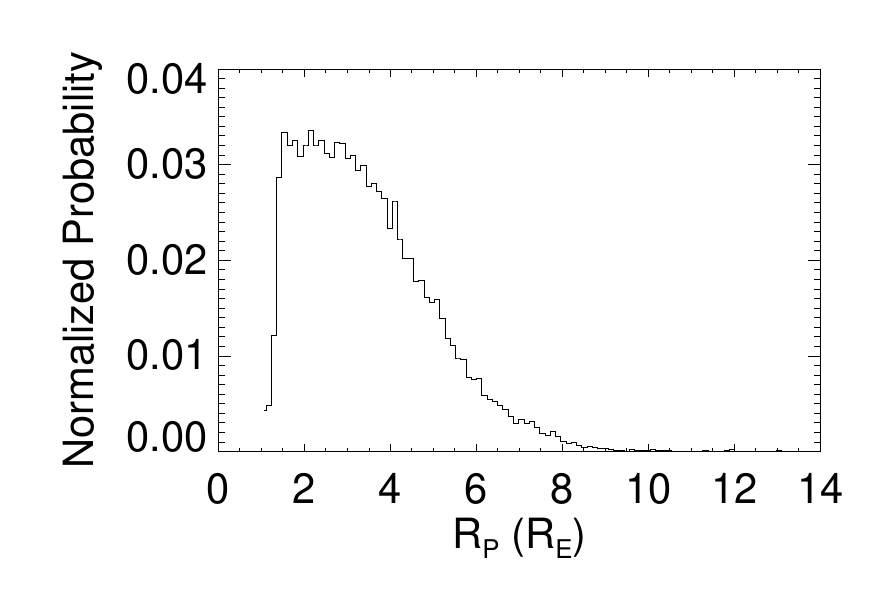}
\caption{The probability distribution function for the radius of planet b. It shows the depth of the transit sets a hard lower limit of $\sim1 \re$. Due to the grazing geometry, the upper envelope is not constrained by the light curve, and instead set by the upper limit on the mass through an RV non-detection and the \citet{Chen:2017} exoplanet mass-radius relation.}
\label{fig:PDF}
\end{figure}

\begin{table}
 \scriptsize
\centering
\setlength\tabcolsep{1.5pt}
\caption{Median values and 68\% confidence intervals for the stellar parameters of the \thisstar and \thisstarcomp from EXOFASTv2 }
  \begin{tabular}{lccccccccccc}
  \hline
\multicolumn{2}{l}{Stellar Parameters}& \thisstar & \thisstarcomp\\
~~~~$M_*$\dotfill &Mass (\msun)\dotfill &$0.686\pm0.033$&$0.581\pm0.029$\\
~~~~$R_*$\dotfill &Radius (\rsun)\dotfill &$0.703^{+0.024}_{-0.022}$&$0.649^{+0.040}_{-0.035}$\\
~~~~$L_*$\dotfill &Luminosity (\lsun)\dotfill &$0.1502\pm0.0057$&$0.0617^{+0.0027}_{-0.0028}$\\
~~~~$\rho_*$\dotfill &Density (cgs)\dotfill &$2.79^{+0.29}_{-0.30}$&$2.99^{+0.56}_{-0.51}$\\
~~~~$\log{g}$\dotfill &Surface gravity (cgs)\dotfill &$4.581^{+0.032}_{-0.037}$&$4.577^{+0.052}_{-0.056}$\\
~~~~$T_{\rm eff}$\dotfill &Effective Temperature (K)\dotfill &$4285^{+49}_{-57}$&$3570\pm110$\\
~~~~$[{\rm Fe/H}]$\dotfill &Metallicity \dotfill &$-0.12^{+0.40}_{-0.42}$&$-0.19^{+0.43}_{-0.81}$\\
~~~~$A_v$\dotfill &V-band extinction \dotfill &$0.029^{+0.018}_{-0.019}$&$0.028^{+0.018}_{-0.019}$\\
~~~~$\sigma_{SED}$\dotfill &SED photometry error scaling \dotfill &$5.0^{+1.6}_{-1.1}$&$5.5^{+1.8}_{-1.2}$\\
~~~~$\pi^{\dag}$\dotfill &Parallax (mas)\dotfill &$12.960\pm0.100$&$12.928\pm0.098$\\
~~~~$d$\dotfill &Distance (pc)\dotfill &$77.16^{+0.60}_{-0.59}$&$77.35^{+0.59}_{-0.58}$\\
  \hline
  \hline
\label{tab:ep248435473_0}
 \end{tabular}
\begin{flushleft}
  \footnotesize $^{\dag}$ The MIST Isochrones were not used in the EXOFASTv2 fit for \thisstar.
  \end{flushleft}
\end{table}

\begin{turnpage}
\begin{table*}
 \scriptsize
\centering
\setlength\tabcolsep{1.5pt}
\caption{Median values and 68\% confidence intervals for planetary parameters of \thisstar from EXOFASTv2 }
  \begin{tabular}{lccccccccccc}
  \hline
  \hline
\multicolumn{2}{l}{Planetary Parameters:}&b&\thisstar.02&c&d&e&\thisstar.06\smallskip\\
~~~~$P$\dotfill &Period (days)\dotfill &$0.658524\pm0.000017$&$6.1002^{+0.0015}_{-0.0017}$&$7.8140^{+0.0019}_{-0.0016}$&$14.69700^{+0.00034}_{-0.00035}$&$19.4820\pm0.0012$&$56.682^{+0.019}_{-0.018}$\\
~~~~$R_P$\dotfill &Radius (\re)\dotfill &$3.3^{+1.8}_{-1.3}$&$0.646^{+0.099}_{-0.091}$&$0.705^{+0.096}_{-0.085}$&$2.93^{+0.14}_{-0.12}$&$2.73^{+0.14}_{-0.11}$&$0.90^{+0.14}_{-0.12}$\\
~~~~$T_C$\dotfill &Time of conjunction (\bjdtdb)\dotfill &$2457945.7235^{+0.0032}_{-0.0030}$&$2457913.413^{+0.013}_{-0.011}$&$2457907.5812^{+0.0099}_{-0.012}$&$2457944.8393\pm0.0012$&$2457938.5410\pm0.0013$&$2457913.436^{+0.014}_{-0.013}$\\
~~~~$T_0$\dotfill &Optimal conjunction Time (\bjdtdb)\dotfill &$2457949.6747^{+0.0032}_{-0.0030}$&$2457943.9143^{+0.0066}_{-0.0064}$&$2457946.6510\pm0.0064$&$2457944.8393\pm0.0012$&$2457938.5410\pm0.0013$&$2457913.436^{+0.014}_{-0.013}$\\
~~~~$a$\dotfill &Semi-major axis (AU)\dotfill &$0.01306^{+0.00020}_{-0.00021}$&$0.05761^{+0.00090}_{-0.00093}$&$0.0679\pm0.0011$&$0.1035^{+0.0016}_{-0.0017}$&$0.1249^{+0.0019}_{-0.0020}$&$0.2546^{+0.0040}_{-0.0041}$\\
~~~~$i$\dotfill &Inclination (Degrees)\dotfill &$75.32^{+0.62}_{-0.70}$&$87.84^{+0.84}_{-0.46}$&$88.28^{+0.81}_{-0.41}$&$89.46^{+0.32}_{-0.25}$&$89.45^{+0.25}_{-0.18}$&$89.40^{+0.26}_{-0.14}$\\
~~~~$e$\dotfill &Eccentricity \dotfill &--&$0.051^{+0.051}_{-0.036}$&$0.042^{+0.043}_{-0.030}$&$0.047^{+0.043}_{-0.032}$&$0.043^{+0.036}_{-0.030}$&$0.31^{+0.11}_{-0.17}$\\
~~~~$\omega_*$\dotfill &Argument of Periastron (Degrees)\dotfill &--&$88^{+60}_{-62}$&$87\pm61$&$87\pm62$&$89^{+57}_{-58}$&$83^{+57}_{-59}$\\
~~~~$T_{eq}$\dotfill &Equilibrium temperature (K)\dotfill &$1515\pm18$&$721.7^{+8.7}_{-8.4}$&$664.5^{+8.0}_{-7.7}$&$538.3^{+6.5}_{-6.3}$&$490.1^{+5.9}_{-5.7}$&$343.3^{+4.1}_{-4.0}$\\
~~~~$M_P$\dotfill &Mass (\me)\dotfill &$11.3^{+11}_{-6.5}$&$0.209^{+0.15}_{-0.089}$&$0.29^{+0.17}_{-0.11}$&$9.4^{+2.9}_{-2.0}$&$8.3^{+2.7}_{-1.8}$&$0.70^{+0.87}_{-0.30}$\\
~~~~$K$\dotfill &RV semi-amplitude (m/s)\dotfill &$10.3^{+10.}_{-5.9}$&$0.094^{+0.067}_{-0.040}$&$0.119^{+0.073}_{-0.046}$&$3.17^{+0.99}_{-0.69}$&$2.53^{+0.84}_{-0.56}$&$0.158^{+0.20}_{-0.068}$\\
~~~~$logK$\dotfill &Log of RV semi-amplitude \dotfill &$1.01^{+0.31}_{-0.37}$&$-1.03^{+0.23}_{-0.24}$&$-0.92\pm0.21$&$0.50^{+0.12}_{-0.11}$&$0.40^{+0.12}_{-0.11}$&$-0.80^{+0.35}_{-0.24}$\\
~~~~$R_P/R_*$\dotfill &Radius of planet in stellar radii \dotfill &$0.043^{+0.023}_{-0.017}$&$0.0084^{+0.0012}_{-0.0011}$&$0.0092^{+0.0012}_{-0.0011}$&$0.03827^{+0.00071}_{-0.00057}$&$0.03564^{+0.00067}_{-0.00061}$&$0.0117^{+0.0018}_{-0.0015}$\\
~~~~$a/R_*$\dotfill &Semi-major axis in stellar radii \dotfill &$4.00^{+0.14}_{-0.15}$&$17.63^{+0.60}_{-0.66}$&$20.80^{+0.71}_{-0.78}$&$31.7^{+1.1}_{-1.2}$&$38.2^{+1.3}_{-1.4}$&$77.9^{+2.7}_{-2.9}$\\
~~~~$\delta$\dotfill &Transit depth (fraction)\dotfill &$0.0018^{+0.0025}_{-0.0012}$&$0.000071^{+0.000022}_{-0.000018}$&$0.000085^{+0.000023}_{-0.000019}$&$0.001465^{+0.000055}_{-0.000043}$&$0.001270^{+0.000048}_{-0.000043}$&$0.000136^{+0.000046}_{-0.000033}$\\
~~~~$Depth$\dotfill &Flux decrement at mid transit \dotfill &$0.000596^{+0.000068}_{-0.000091}$&$0.000071^{+0.000022}_{-0.000018}$&$0.000085^{+0.000023}_{-0.000019}$&$0.001465^{+0.000055}_{-0.000043}$&$0.001270^{+0.000048}_{-0.000043}$&$0.000136^{+0.000046}_{-0.000033}$\\
~~~~$\tau$\dotfill &Ingress/egress transit duration (days)\dotfill &$0.00686^{+0.00051}_{-0.00045}$&$0.00116^{+0.00042}_{-0.00032}$&$0.00134^{+0.00045}_{-0.00033}$&$0.00569^{+0.00067}_{-0.00045}$&$0.00601^{+0.00073}_{-0.00063}$&$0.00281^{+0.0019}_{-0.00096}$\\
~~~~$T_{14}$\dotfill &Total transit duration (days)\dotfill &$0.01389^{+0.0012}_{-0.00085}$&$0.083^{+0.014}_{-0.012}$&$0.094^{+0.016}_{-0.014}$&$0.1420^{+0.0014}_{-0.0015}$&$0.1527^{+0.0014}_{-0.0015}$&$0.143^{+0.024}_{-0.023}$\\
~~~~$T_{FWHM}$\dotfill &FWHM transit duration (days)\dotfill &$0.00695^{+0.00062}_{-0.00042}$&$0.082^{+0.015}_{-0.013}$&$0.092^{+0.016}_{-0.014}$&$0.1362\pm0.0015$&$0.1466^{+0.0015}_{-0.0016}$&$0.140\pm0.024$\\
~~~~$b$\dotfill &Transit Impact parameter \dotfill &$1.011^{+0.027}_{-0.024}$&$0.64^{+0.13}_{-0.25}$&$0.60^{+0.14}_{-0.28}$&$0.29^{+0.12}_{-0.17}$&$0.36^{+0.10}_{-0.16}$&$0.64^{+0.17}_{-0.33}$\\
~~~~$b_S$\dotfill &Eclipse impact parameter \dotfill &--&$0.68^{+0.13}_{-0.26}$&$0.64^{+0.14}_{-0.30}$&$0.31^{+0.12}_{-0.18}$&$0.379^{+0.100}_{-0.16}$&$0.86^{+0.30}_{-0.33}$\\
~~~~$\tau_S$\dotfill &Ingress/egress eclipse duration (days)\dotfill &--&$0.00131^{+0.00055}_{-0.00038}$&$0.00148^{+0.00057}_{-0.00039}$&$0.00613^{+0.00064}_{-0.00048}$&$0.00641^{+0.00069}_{-0.00052}$&$0.0037^{+0.0035}_{-0.0037}$\\
~~~~$T_{S,14}$\dotfill &Total eclipse duration (days)\dotfill &--&$0.084^{+0.019}_{-0.016}$&$0.096^{+0.020}_{-0.017}$&$0.1485^{+0.0100}_{-0.0051}$&$0.1593^{+0.010}_{-0.0052}$&$0.13^{+0.10}_{-0.13}$\\
~~~~$T_{S,FWHM}$\dotfill &FWHM eclipse duration (days)\dotfill &--&$0.083^{+0.020}_{-0.017}$&$0.094^{+0.021}_{-0.018}$&$0.1424^{+0.0099}_{-0.0052}$&$0.1528^{+0.011}_{-0.0053}$&$0.13^{+0.11}_{-0.13}$\\
~~~~$\delta_{S,3.6\mu m}$\dotfill &Blackbody eclipse depth at 3.6$\mu$m (ppm)\dotfill &$210^{+290}_{-140}$&$0.41^{+0.14}_{-0.11}$&$0.299^{+0.088}_{-0.069}$&$1.24^{+0.15}_{-0.12}$&$0.511^{+0.070}_{-0.057}$&$0.00160^{+0.00062}_{-0.00044}$\\
~~~~$\delta_{S,4.5\mu m}$\dotfill &Blackbody eclipse depth at 4.5$\mu$m (ppm)\dotfill &$280^{+380}_{-180}$&$0.94^{+0.31}_{-0.24}$&$0.76^{+0.22}_{-0.17}$&$4.25^{+0.44}_{-0.35}$&$2.05^{+0.24}_{-0.19}$&$0.0135^{+0.0050}_{-0.0036}$\\
~~~~$\rho_P$\dotfill &Density (cgs)\dotfill &$1.77^{+1.9}_{-0.88}$&$4.27^{+0.79}_{-0.66}$&$4.51^{+0.81}_{-0.67}$&$2.03^{+0.64}_{-0.43}$&$2.21^{+0.70}_{-0.47}$&$5.27^{+1.5}_{-0.87}$\\
~~~~$logg_P$\dotfill &Surface gravity \dotfill &$3.00^{+0.14}_{-0.13}$&$2.69\pm0.12$&$2.75\pm0.11$&$3.026^{+0.12}_{-0.100}$&$3.03^{+0.12}_{-0.10}$&$2.93^{+0.18}_{-0.12}$\\
~~~~$\Theta$\dotfill &Safronov Number \dotfill &$0.0046^{+0.0018}_{-0.0015}$&$0.00191^{+0.00094}_{-0.00065}$&$0.00285^{+0.0012}_{-0.00086}$&$0.0340^{+0.010}_{-0.0072}$&$0.0387^{+0.012}_{-0.0082}$&$0.0203^{+0.017}_{-0.0070}$\\
~~~~$\fave$\dotfill &Incident Flux (\fluxcgs)\dotfill &$1.197^{+0.059}_{-0.055}$&$0.0612^{+0.0031}_{-0.0028}$&$0.0441^{+0.0022}_{-0.0020}$&$0.01898^{+0.00093}_{-0.00086}$&$0.01304^{+0.00064}_{-0.00060}$&$0.00286^{+0.00025}_{-0.00024}$\\
~~~~$T_P$\dotfill &Time of Periastron (\bjdtdb)\dotfill &$2457945.7235^{+0.0032}_{-0.0030}$&$2457913.43^{+0.93}_{-0.95}$&$2457907.6\pm1.2$&$2457944.8\pm2.3$&$2457938.6\pm2.9$&$2457913.3^{+5.0}_{-5.2}$\\
~~~~$T_S$\dotfill &Time of eclipse (\bjdtdb)\dotfill &$2457946.0528^{+0.0032}_{-0.0030}$&$2457916.46\pm0.17$&$2457903.67^{+0.19}_{-0.18}$&$2457937.49^{+0.41}_{-0.40}$&$2457948.28\pm0.44$&$2457885.2^{+8.6}_{-8.7}$\\
~~~~$T_A$\dotfill &Time of Ascending Node (\bjdtdb)\dotfill &$2457945.5589^{+0.0032}_{-0.0030}$&$2457911.942^{+0.13}_{-0.076}$&$2457905.682^{+0.14}_{-0.080}$&$2457941.28^{+0.28}_{-0.17}$&$2457933.82^{+0.32}_{-0.19}$&$2457903.3^{+3.4}_{-4.2}$\\
~~~~$T_D$\dotfill &Time of Descending Node (\bjdtdb)\dotfill &$2457945.8881^{+0.0032}_{-0.0030}$&$2457914.884^{+0.077}_{-0.13}$&$2457909.478^{+0.078}_{-0.14}$&$2457948.40^{+0.17}_{-0.28}$&$2457943.26^{+0.19}_{-0.33}$&$2457923.6^{+4.3}_{-3.5}$\\
~~~~$ecos{\omega_*}$\dotfill & \dotfill &--&$-0.000^{+0.044}_{-0.045}$&$0.000\pm0.037$&$-0.000\pm0.043$&$-0.000^{+0.036}_{-0.035}$&$0.00\pm0.24$\\
~~~~$esin{\omega_*}$\dotfill & \dotfill &--&$0.026^{+0.046}_{-0.022}$&$0.022^{+0.038}_{-0.018}$&$0.025^{+0.035}_{-0.020}$&$0.025^{+0.034}_{-0.020}$&$0.19^{+0.16}_{-0.15}$\\
~~~~$V/V_C$\dotfill & \dotfill &--&$0.973^{+0.022}_{-0.044}$&$0.978^{+0.018}_{-0.037}$&$0.975^{+0.020}_{-0.034}$&$0.975^{+0.020}_{-0.033}$&$0.80^{+0.15}_{-0.12}$\\
~~~~$M_P\sin i$\dotfill &Minimum mass (\me)\dotfill &$10.9^{+11}_{-6.3}$&$0.208^{+0.15}_{-0.089}$&$0.29^{+0.17}_{-0.11}$&$9.4^{+2.9}_{-2.0}$&$8.3^{+2.7}_{-1.8}$&$0.70^{+0.87}_{-0.30}$\\
~~~~$M_P/M_*$\dotfill &Mass ratio \dotfill &$0.000049^{+0.000051}_{-0.000028}$&$0.00000091^{+0.00000065}_{-0.00000039}$&$0.00000126^{+0.00000077}_{-0.00000049}$&$0.0000413^{+0.000013}_{-0.0000091}$&$0.0000362^{+0.000012}_{-0.0000081}$&$0.0000031^{+0.0000039}_{-0.0000013}$\\
~~~~$d/R_*$\dotfill &Separation at mid transit \dotfill &$4.00^{+0.14}_{-0.15}$&$16.98^{+0.81}_{-0.92}$&$20.17^{+0.87}_{-1.0}$&$30.7^{+1.4}_{-1.7}$&$37.0^{+1.6}_{-1.7}$&$59^{+13}_{-11}$\\
~~~~$P_T$\dotfill &A priori non-grazing transit prob \dotfill &$0.2389^{+0.010}_{-0.0092}$&$0.0584^{+0.0033}_{-0.0027}$&$0.0491^{+0.0026}_{-0.0020}$&$0.0313^{+0.0018}_{-0.0014}$&$0.0261^{+0.0012}_{-0.0011}$&$0.0166^{+0.0039}_{-0.0030}$\\
~~~~$P_{T,G}$\dotfill &A priori transit prob \dotfill &$0.2616^{+0.011}_{-0.0098}$&$0.0594^{+0.0034}_{-0.0027}$&$0.0500^{+0.0026}_{-0.0021}$&$0.0338^{+0.0020}_{-0.0015}$&$0.0280^{+0.0013}_{-0.0012}$&$0.0170^{+0.0039}_{-0.0030}$\\
~~~~$P_S$\dotfill &A priori non-grazing eclipse prob \dotfill &--&$0.0546\pm0.0025$&$0.0465\pm0.0020$&$0.02941^{+0.0013}_{-0.00093}$&$0.0245^{+0.0012}_{-0.0010}$&$0.0113^{+0.0017}_{-0.0016}$\\
~~~~$P_{S,G}$\dotfill &A priori eclipse prob \dotfill &--&$0.0555\pm0.0026$&$0.0473^{+0.0021}_{-0.0020}$&$0.0317^{+0.0015}_{-0.0010}$&$0.0263^{+0.0013}_{-0.0012}$&$0.0116^{+0.0018}_{-0.0017}$\\\label{tab:ep248435473_1}
 \end{tabular}
\end{table*}
\end{turnpage}
\begin{table}
\label{tab:other}
 \tiny
\centering
\setlength\tabcolsep{1.0pt}
\caption{Median values and 68\% confidence intervals for the additional parameters of \thisstar from EXOFASTv2 }
  \begin{tabular}{lccccccccccc}
  \hline
  \hline
\multicolumn{2}{l}{Wavelength Parameters:}&Kepler\smallskip\\
~~~~$u_{1}$\dotfill &linear limb-darkening coeff \dotfill &$0.630^{+0.064}_{-0.10}$\\
~~~~$u_{2}$\dotfill &quadratic limb-darkening coeff \dotfill &$0.107^{+0.082}_{-0.054}$\\
\multicolumn{2}{l}{Telescope Parameters:}&TRES\smallskip\\
~~~~$\gamma_{\rm rel}$\dotfill &Relative RV Offset (m/s)\dotfill &$-29.9^{+7.4}_{-8.6}$\\
~~~~$\sigma_J$\dotfill &RV Jitter (m/s)\dotfill &$0.00^{+19}_{-0.00}$\\
~~~~$\sigma_J^2$\dotfill &RV Jitter Variance \dotfill &$-130^{+480}_{-140}$\\
\multicolumn{2}{l}{Transit Parameters:}&\\
Planet& Transit Date &Added Variance \dotfill&Transit Mid Time\dotfill &Baseline flux \dotfill\\
&&~~~~$\sigma^{2}\times 10^{-10}$\dotfill &$\bjdtdb$\dotfill &~~~~$F_0$\dotfill \\
b,c,d,g &       Full K2 LC & $7.6^{+2.9}_{-2.7}$ &                          N/A & $1.0000035\pm0.0000030$            \\
e       & K2 UT 2017-06-10 &    $19^{+26}_{-18}$ & $2457915.44761 \pm 0.00106 $ & $0.999981\pm0.000016$              \\
f       & K2 UT 2017-06-14 &    $10^{+21}_{-15}$ & $2457919.05628 \pm 0.00088 $ & $1.000012\pm0.000014$              \\
e       & K2 UT 2017-06-25 &    $-3^{+15}_{-11}$ & $2457930.13813 \pm 0.00101 $ & $1.000002^{+0.000013}_{-0.000012}$ \\
f       & K2 UT 2017-07-04 &    $25^{+26}_{-18}$ & $2457938.54211 \pm 0.00108 $ & $0.999986^{+0.000016}_{-0.000017}$ \\
e       & K2 UT 2017-07-10 &    $16^{+25}_{-17}$ & $2457944.83597 \pm 0.00137 $ & $0.999971^{+0.000016}_{-0.000015}$ \\
f       & K2 UT 2017-07-23 &    $54^{+41}_{-27}$ & $2457958.03343 \pm 0.00112 $ & $0.999986^{+0.000020}_{-0.000021}$ \\
e       & K2 UT 2017-07-25 &    $13^{+21}_{-14}$ & $2457959.52590 \pm 0.00090 $ & $1.000000^{+0.000014}_{-0.000015}$ \\
e       & K2 UT 2017-08-08 &    $30^{+26}_{-19}$ & $2457974.23919 \pm 0.00114 $ & $1.000009\pm0.000017$              \\
f       & K2 UT 2017-08-12 &     $5^{+23}_{-15}$ & $2457977.50614 \pm 0.00120 $ & $1.000007\pm0.000015$              \\

\smallskip\\
  \hline
  \hline
\label{tab:ep248435473_2}
 \end{tabular}
\end{table}

\begin{table*}
 \centering
 \caption{The Best Confirmed or Validated Planets for Transmission Spectroscopy with R$_P$ $<$ 3\rearth}
 \label{tbl:S/N}
 \begin{tabular}{cccccc}
    \hline
    \hline
    Planet & Period (days) & R$_P$(\rearth) &  S/N$^a$ & Reference & Discovery \\
    \hline
GJ 1214 b  &1.58& 2.85 & 1.00 & \citet{Charbonneau:2009} & MEarth\\
 55 Cnc e  &0.74& 1.91 & 0.41 & \citet{Dawson:2010} & RVs\\
 HD 97658 b  & 9.49 & 2.35 & 0.27 & \citet{Dragomir:2013} & RVs\\
 TRAPPIST-1 f  &9.21& 1.04 & 0.24 & \citet{Gillon:2017} & \textit{Spitzer}\\
 K2-136 c  &17.31& 2.91 & 0.19 & \citet{Ciardi:2018, Livingston:2018,Mann:2018} & K2\\
 GJ 9827 b  &1.21& 1.75 & 0.18 & \citet{Niraula:2017, Rodriguez:2018} & K2\\
 K2-167 b  &9.98&2.82 & 0.16 & \citet{Mayo:2018} & K2\\
\textbf{K2-266 e} &\textbf{14.70}& \textbf{2.93} & \textbf{0.15} & \textbf{This Work} & \textbf{K2}\\
 GJ 9827 d  &6.20&2.10 & 0.15 & \citet{Niraula:2017, Rodriguez:2018} & K2\\
 HIP 41378 b  &15.57& 2.90 & 0.14 & \citet{Vanderburg:2016b} & K2\\
 HD 3167 b  &0.96& 1.70 & 0.14 & \citet{Vanderburg:2016c, Christiansen:2017} & K2\\
 K2-233 d  &24.37& 2.65 & 0.13 & \citet{David:2018} & K2\\
 \textbf{K2-266 f} &\textbf{19.48}& \textbf{2.73} & \textbf{0.12} & \textbf{This Work} & \textbf{K2}\\
K2-28 b  &2.26& 2.32 & 0.12 & \citet{Hirano:2016} & K2\\
 K2-199 c  &7.37& 2.84 & 0.12 &\citet{Mayo:2018} & K2 \\
 K2-155 c  &13.85& 2.60 & 0.11 & \citet{Diez:2018, Hirano:2018} & K2\\
 Kepler-410 A b  &17.83& 2.84 & 0.10 & \citet{VanEylen:2014} & \Kepler\\
 HD 106315 b  &9.55& 2.40 & 0.10 & \citet{Rodriguez:2017b, Crossfield:2017}  & K2\\
   \hline
    \hline
 \end{tabular}
\begin{flushleft} 
 \footnotesize{ \textbf{\textsc{NOTES:}}
$^a$The predicted signal-to-noise ratios relative to GJ 1214 b. All values used in determining the signal-to-noise were obtained from the NASA Exoplanet Archive \citep{Akeson:2013}. If a system did not have a reported mass on NASA Exoplanet Archive or it was not a 2$\sigma$ result, we used the \citet{Weiss:2014} Mass-Radius relationship to estimate the planet's mass. $^b$Our calculation for the S/N of 55 Cnc e assumes a H/He envelope since it falls just above the pure rock line determined by \citet{Zeng:2016}. However, 55 Cnc e is in a ultra short period orbit, making it unlikely that it would hold onto a thick H/He envelope. We do not include \thisstar\ b due to its grazing configuration. \\
}
\end{flushleft}
\end{table*}

\section{Dynamics of \thisstar} 
\label{sec:dynamics} 

Given its multiplicity and mutually-transiting nature, the six-planet system orbiting \thisstar\ can be classified as one of the Systems of Tightly Packed Inner Planets (STIPs) common in the \Kepler\ data \citep{Lissauer:2011,VanLaerhoven:2012,Swift:2013}. However, this system is unique due to the innermost planetary orbit displaying a remarkable 75 degree inclination and a grazing transit. Members of the \Kepler\ multi-planet systems have smaller mutual inclinations, typically within a few degrees of each other \citep{Fang:2012,Figueira:2012, Fabrycky:2014}. Moreover, these systems do not tend to excite high mutual inclinations without some external factor \citep{Becker:2016,Mustill:2017, Hansen:2017,Becker:2017,Jontof-Hutter:2017, Denham:2018}. In this section, we discuss the information gained through combining the observed light curve with dynamical analysis, and attempt to constrain the current dynamical state of the system. 

\subsection{Transit Timing Variations} 
The transit timing measurements for planets d and e listed in Table \ref{tab:other} can be used to derive dynamical constraints on their masses and orbits. In this section, we invert the planet pairs' TTVs to infer their masses and, in combination with the planet radii derive from out light-curve fitting, their densities.
We model the planets'  TTVs using the TTVFast code developed by \citet{Deck:2014} and use the {\texttt emcee}  package's \citep{ForemanMackey:2012} ensemble sampler, based on the algorithm of \citet{GoodmanWeare2010}, to sample the posterior distribution of the planetary masses and orbital elements. We model only the dynamical interactions of planets d and e, and ignore any perturbations from the other (potential) members of the system.\footnote{The variations induced by the additional planets in the system are expected to be negligible. For example, assuming a $1~M_\oplus$ planet c and circular orbits, it induces variations of less than $10$ seconds in planet d's transit times. Allowing for modest eccentricities does not significantly enhance the induced TTVs. The influence of the other additional planets is expected to be even weaker given as they are more widely separated and do not fall near any significant resonances with d or e.} We assume planets d and e orbit in the same plane since small mutual inclinations have negligible influence on TTVs.  We approximate the mid-transit uncertainties to be  Gaussian-distributed about the median transit time determined by EXOFASTv2, with variances set to the larger of the two asymmetric error bars in Table \ref{tab:ep248435473_2}. Our likelihood function is then computed based on the standard chi-squared statistic as $\ln{\cal L} = -\frac{1}{2}\chi^2$. We impose a Gaussian prior with 0 mean and a variance of $\sigma=0.05$ on the eccentricity vector components, $e_i\cos\omega_i$ and $e_i\sin\omega_i$, typical for eccentricities of multi-planet, sub-Neptune systems \citep{Hadden:2014,Hadden:2017,VanEylen2015,Xie2016}
We found that the inferred planet masses are insensitive to the assumed eccentricity priors after running additional MCMC fits with $\sigma=0.025$ and $\sigma=0.1$.
We initialize our MCMC with 200 walkers and run for 120,000 iterations, saving all walker positions every 1000 iterations.
    
\begin{figure}
     \centering
     \includegraphics[width=\columnwidth]{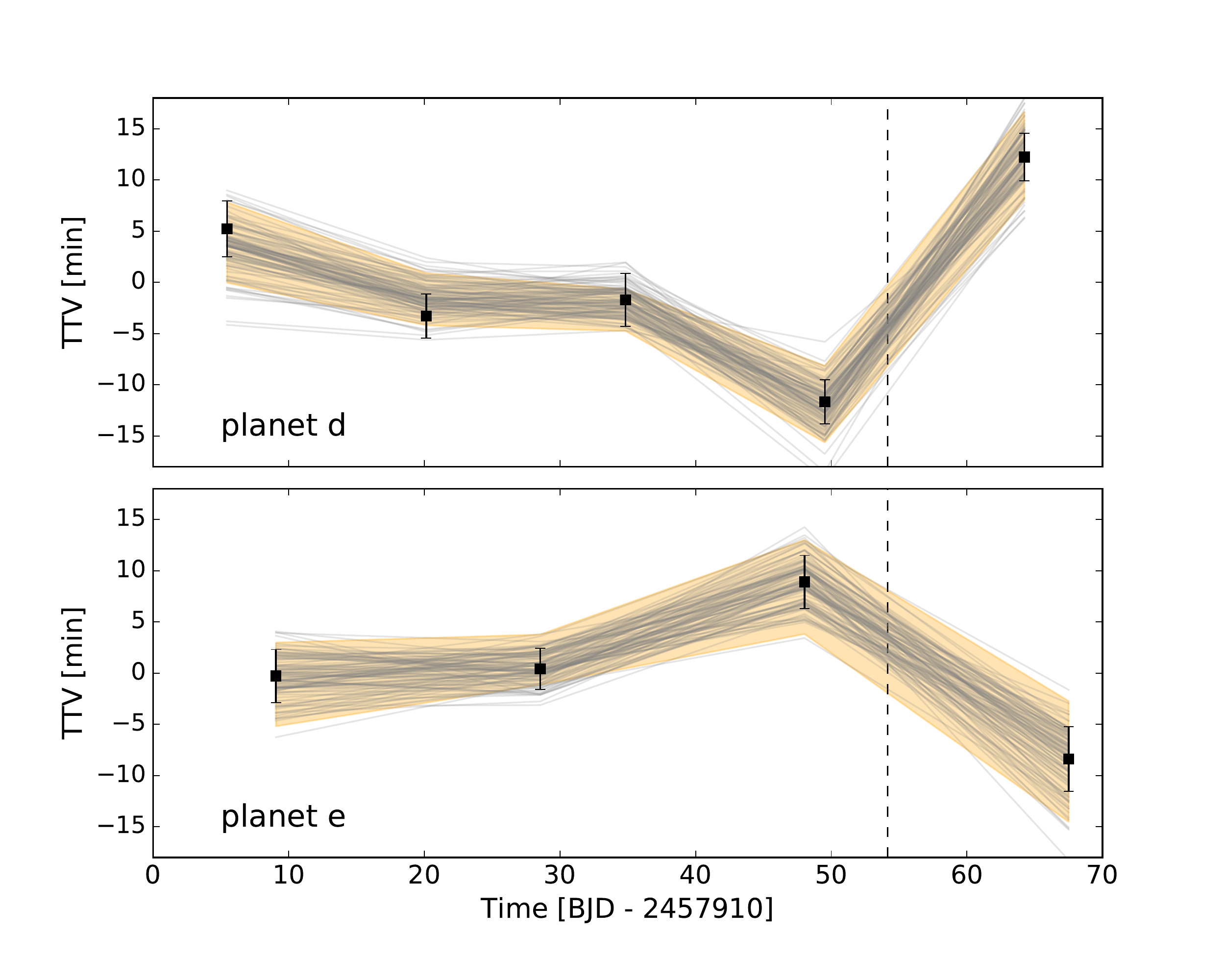}
     \caption{The observed transit timing variations of planet's d and e are shown by black squares with error bars showing the $1\sigma$ uncertainties. A sample of 100 $N$-body solutions drawn from our MCMC fit posteriors are shown in gray. The shaded orange region shows the 1$\sigma$ uncertainty in the analytic TTV model fit via least-squares. The time at which d and e experience a conjunction is indicated by the dashed vertical.}
     \label{fig:ef_ttv}
 \end{figure}

\begin{figure*}
     \centering
     \includegraphics[width=0.8\textwidth]{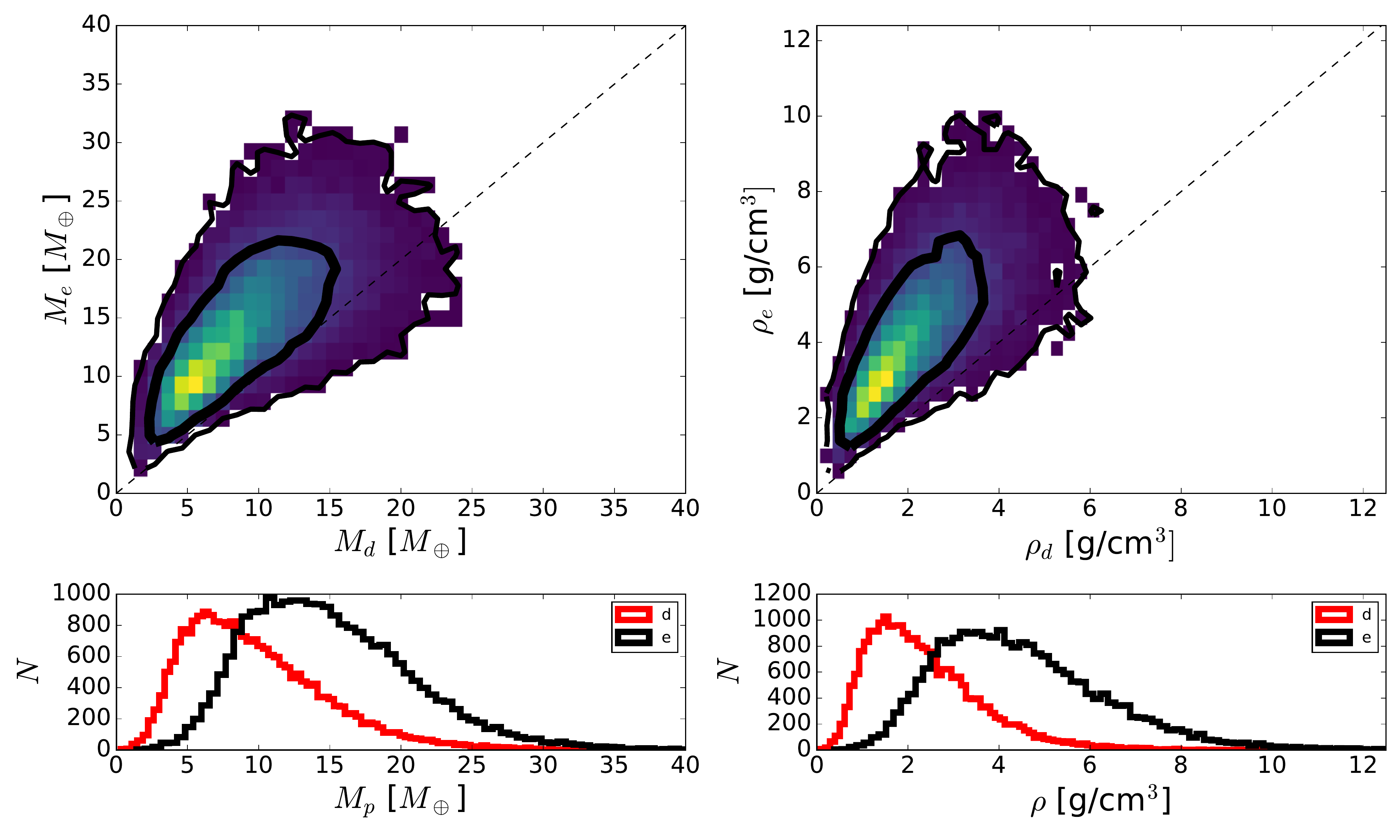}
     \caption{Posterior distributions of planet d and e's masses and densities derived from TTV dynamical modeling. Joint posterior distributions are plotted as intensity maps in the top two panels with 68\% and 95\% credible region contours plotted in black. Equal mass and density  are indicated with dashed lines. Bottom panels show one dimensional histograms of the marginalized posterior distribution of planet d (red) and e's (back) mass and density.}
     \label{fig:ef_ttv_post}
 \end{figure*}

Figure \ref{fig:ef_ttv} shows the observed timing variations of planets d and e, along with $N$-body solutions drawn from our MCMC posterior. From our TTV dynamical fit, we determine planet-star mass ratios of $m_d/M_* = 3.9^{+2.5}_{-1.7}\times10^{-5}$ for planet d and $m_e/M_* =6.3^{+2.8}_{-2.2}\times10^{-5}$ for planet e.  The TTVs yield no strong constraint on planet eccentricities and the posterior distributions largely mirror our assumed priors. We convert the  dynamical constraints on planet-star mass ratios to constraints on the planetary masses and densities by combining the posterior samples from our TTV fit with posterior samples of our fit to stellar mass and planet radii computed with EXOFASTv2. The resulting posterior distributions of the planets' masses and densities are plotted in Figure \ref{fig:ef_ttv_post}. The inferred median and $1\sigma$ planet mass values are $m_d= 8.4_{-3.6}^{+5.4} M_\oplus$ and $m_e=13.6_{-4.7}^{+6.1} M_\oplus$ and densities are $\rho_d= 2.7_{-1.2}^{+1.8}~\text{g/cm}^3$ and
$\rho_e=5.6_{-2.0}^{+2.6}~\text{g/cm}^3$.
    
    Our $N$-body dynamical model contains 10 free parameters which are fit to only nine data points. This means that, at face value, our model is under-constrained and we are at risk of over-fitting. To understand the origin of our dynamical mass constraints and ensure that they are not merely the result of over-fitting, we analyze the TTVs using the analytic model of \citet{Hadden:2016}. This analytic treatment reduces the dimensionality of the TTV model so that it is no longer under-constrained. Note that we adopt the masses derived from the more complete $N$-body model as our best fit values; we present the analytic model simply as a consistency check to ensure the $N$-body results are not over-fitting the data because of poor MCMC convergence.
    
    We use the formulas of \citet{Hadden:2016} to construct an analytic model for the TTVs of planets d and e as a function of planet periods $P_i$, initial transit times $T_i$, planet-star mass ratios $\mu_i$, and the complex `combined eccentricity' 
    \begin{equation}
        {\cal Z}\approx\frac{1}{\sqrt{2}}( e_ee^{i\omega_e}-e_d e^{i\omega_d})~.
    \end{equation} 
    The analytic model reduces the total number of model parameters to 8 by combining the planets' eccentricities and longitudes of perihelia into the single complex quantity, ${\cal Z}$.
    The $n$th transit of planet d and e are modeled as 
    \begin{eqnarray}
        t_i(n) = T_{i} + n P_i + \delta t_{{\cal C},i}(n) + \delta t_{{\cal F},i}(n)
        \label{eq:ttv_e}
    \end{eqnarray}
    where  $\delta t_{\cal C}\propto \mu'$, $\delta t_{\cal F} \propto \mu'{\cal Z}$ with $\mu'$ the perturbing planet's planet-star mass ratio. Expressions for $\delta t_{\cal C}$ and $\delta t_{\cal F}$ are given in \citet{Hadden:2016}.
    We use the  Levenberg-Marquardt minimization algorithm to fit our analytic model to the observed transit times and estimate uncertainties from the local curvature of the $\chi^2$ surface  \citep[e.g.,][]{Press:1992}. 
    The analytic fit, plotted in Figure \ref{fig:ef_ttv}, yields masses $m_d/M_* = 3.0\pm0.8\times10^{-5}$ for planet d and $m_e/M_* =4.6\pm1.0\times10^{-5}$ for planet e, which are consistent with the N-body MCMC constraints. 
    
    The origins of the mass constraint can be qualitatively understood from the analytic model as follows: at conjunction, planets impart impulsive kicks to one another that change their instantaneous orbital periods. This effect is captured by the so-called `chopping' terms, $\delta t_{\cal C}$, in Equation \eqref{eq:ttv_e} \citep[see also][]{Nesvorny:2014,Deck2015}.
    Indeed, we obtain nearly identical mass constraints from an analytic fit that drops the $\delta t_{\cal F}$ terms from Equation \eqref{eq:ttv_e} (and thereby further reduces the number of free parameters to 6 as the model no longer depends on the complex number ${\cal Z}$). Because these $\delta t_{\cal F}$ terms vary over a timescale much longer than the baseline of our observations, they are well-approximated by a linear trend and essentially degenerate with small changes to the $T_{i} + n P_i$ terms in Equation \eqref{eq:ttv_e}. Thus, we have identified the the origin of our mass constraints with the measurement of the chopping signals, $\delta t_{{\cal C},i}$, in d and e's TTVs.
    Over the course of our observing baseline, planet d and e experience a single conjunction, at the time marked by a dashed line in Figure \ref{fig:ef_ttv}. The power of the TTV signal for constraining the planets' masses comes mainly from the impulsive changes in the planets' osculating orbital periods experienced at this conjunction causing the planets to arrive early (in the case of e) or late (in the case of d) at their next transits.

\subsection{Dynamical Stability and Transit Likelihood} 

Next we consider the dynamical stability of the system, along with the probability that all of the putative planets can be seen in transit. Although most Kepler multi-planet systems tend to have low mutual inclinations, this system is unique to date because there is a significant mutual inclination between the innermost planet and the other five. In the context of the known set of multi-planet systems, this appears significant.
\citet{Ballard:2016} found that the Kepler planet-hosting systems around cool stars appears to be drawn from two populations: a set of multi-transit systems and a second set of single-transit planets, which may also have statistically higher obliquities \cite{Morton:2014a} (this concept of these two populations is commonly called the `Kepler dichotomy').
One solution to the Kepler dichotomy is that the two populations are actually all multi-planet systems, and that the ones that appear to be singly-transiting are systems with larger mutual inclinations, which can see seen as single-transit systems from a particular line of sight.
Although most Kepler multi-planet (4 planets or more) are fairly tightly confined to a roughly coplanar region, there is some precedent for multi-planet systems: \citet{Mills:2017} found a two-planet system with a 24 degree mutual inclination. 
In cases like this, the question of how many planets in a multi-transiting system might be seen in transit at any one time becomes relevant \citep{Brakensiek:2016}, as large mutual inclinations can lead to only a subset of the planets being seen from a given line of sight. 
\thisstar b is currently observed to have a grazing transit and a high mutual inclination with the remainder of the planets, the five of which reside in a roughly coplanar configuration.
With an aim towards assessing where this system fits into the Kepler dichotomy, we in this section conduct an analysis of the transit likelihood for various numbers of planets in this system.

\begin{figure}
     \centering
     \includegraphics[width=0.45\textwidth]{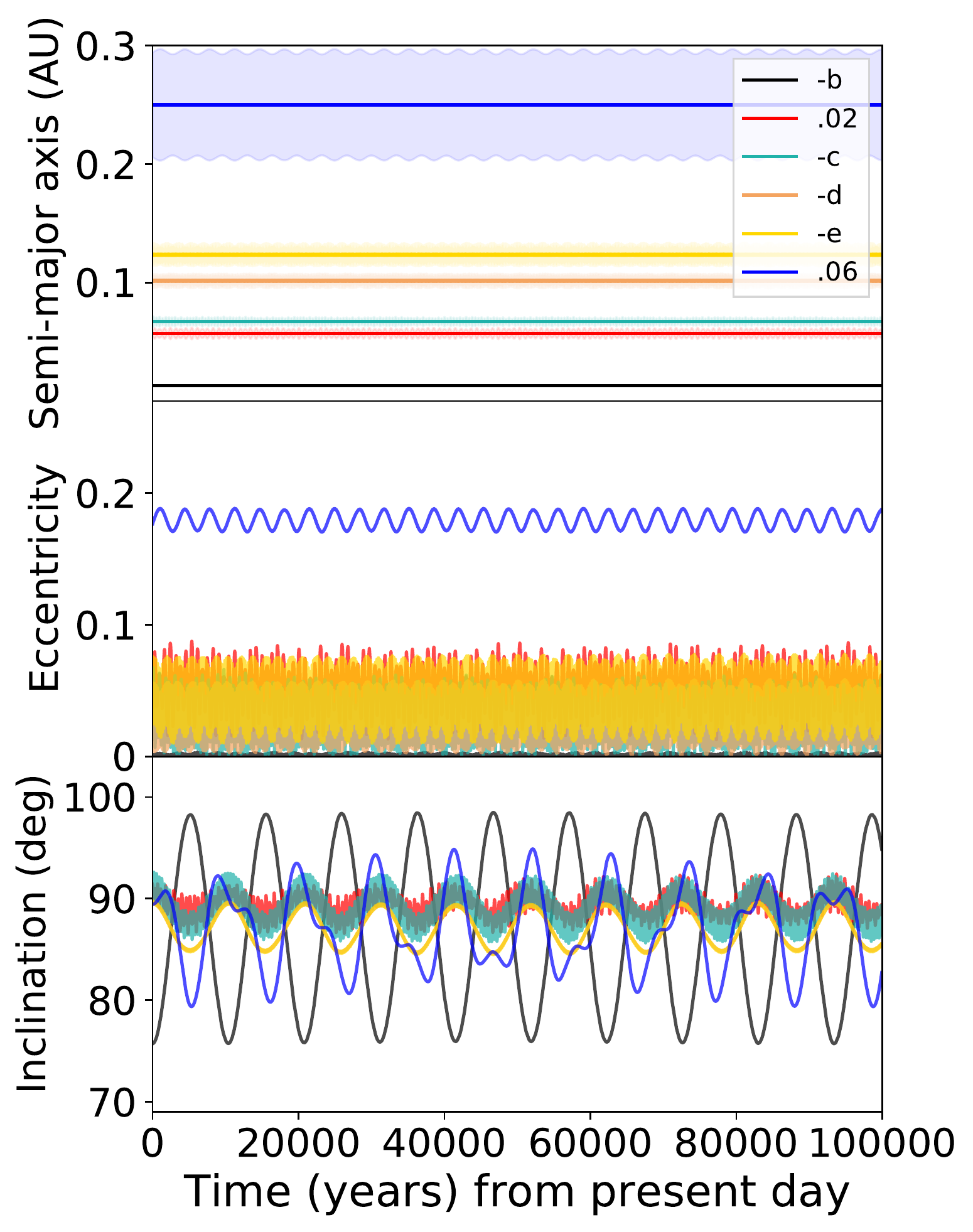}
     \caption{A sample integration from the suite of 250 run for this work, plotted for illustrative purposes. A typical dynamically stable integration, where planetary orbits do not wander far from their initial values of semi-major axis or eccentricity. (top panel) The semi-major axis of each planet, with shaded regions to denote the extent of the perihelion and aphelion distances. (middle panel) The eccentricity evolution of each planet, which oscillate but remain confined near their initial values. (bottom panel) Inclination values for each planet. The inclination values for all planets oscillate with varying amplitudes, as is typical for all integrations.  }
     \label{fig:num_sims_compare}
 \end{figure}

To test the dynamical and transit stability of these planets, we ran 250 N-body simulations of the system, drawing the initial orbital elements from the posteriors generated from the EXOFAST transit fit (more specifically, we draw a single link from the MCMC posterior at random for each of the 250 simulations, and then use all orbital elements from that link). 
We assigned the longitude of ascending node to be $2\pi$, as it cannot be measured from the transit fits. 
The planetary masses are drawn from the posteriors provided by the EXOFAST fit, as are stellar mass and radius. 
For all calculated values of inclination, we broke the above/below solar mid-plane degeneracy by randomly assigning the value to be either greater or smaller than 90 degrees. We also assume that the stellar obliquity is aligned with the plane containing the outer five planets (but there are no system-specific observations to support this assumption; instead, we make this assumption as a computational necessity, although we expect the stellar obliquity to be more aligned in multi-planet systems; \citealt{Morton:2014a}). These 250 simulations were carried out using the hybrid Wisdom-Holman and Bulirsch-Stoer (B-S) integrator \texttt{Mercury6} \citep{Chambers:1999} for integration times of 10$^{5}$ years, and with an initial time-step of 8.5 minutes. Energy was conserved to better than one part in  $10^{8}$ over the course of the simulations. When physical collisions occur, particles are removed from the simulation. The integration time was chosen to include many secular timescales of the system (Figure \ref{fig:num_sims_compare} shows that many periods of secular oscillations are included in 10$^{5}$ years time span).
 
In roughly 66\% of our suite of 250 simulations, at least two planets in the system attain orbits which cross. In 23\% of the simulations, the system experiences a true dynamical instability, in which a planet is ejected from the system or physically collides with another body. In the cases in which orbits cross, planets 0.02 and c are the culprits of the instability roughly 80\% of the time. On the $10^{5}$ year integrations considered in this work, a size-able fraction (roughly a third) remain dynamically stable. As such, we cannot use dynamical arguments to argue against the existence of planet candidate 0.02, whose close orbit with planet c might otherwise be suspect.

As neither of the planet candidates can clearly be ruled out based on dynamical arguments, we next consider the dynamical evolution that leads to only a subset of the six known planets being seen in transit. 
We currently observe the system to be a six-planet system, but the innermost planet \thisstar b has a high measured impact parameter, meaning that is barely transiting. The simulations show significant inclination evolution over time for both \thisstar b and the other planets in the system. In Figure \ref{fig:num_sims_compare}, we plot a representative case from our set of 250 simulations, where the semi-major axes and eccentricities of all planets remain confined relatively close to their currently-measured values, but the orbital inclination of all six planets evolves.

\begin{figure}[!ht] 
\centering 
\includegraphics[width=\columnwidth]{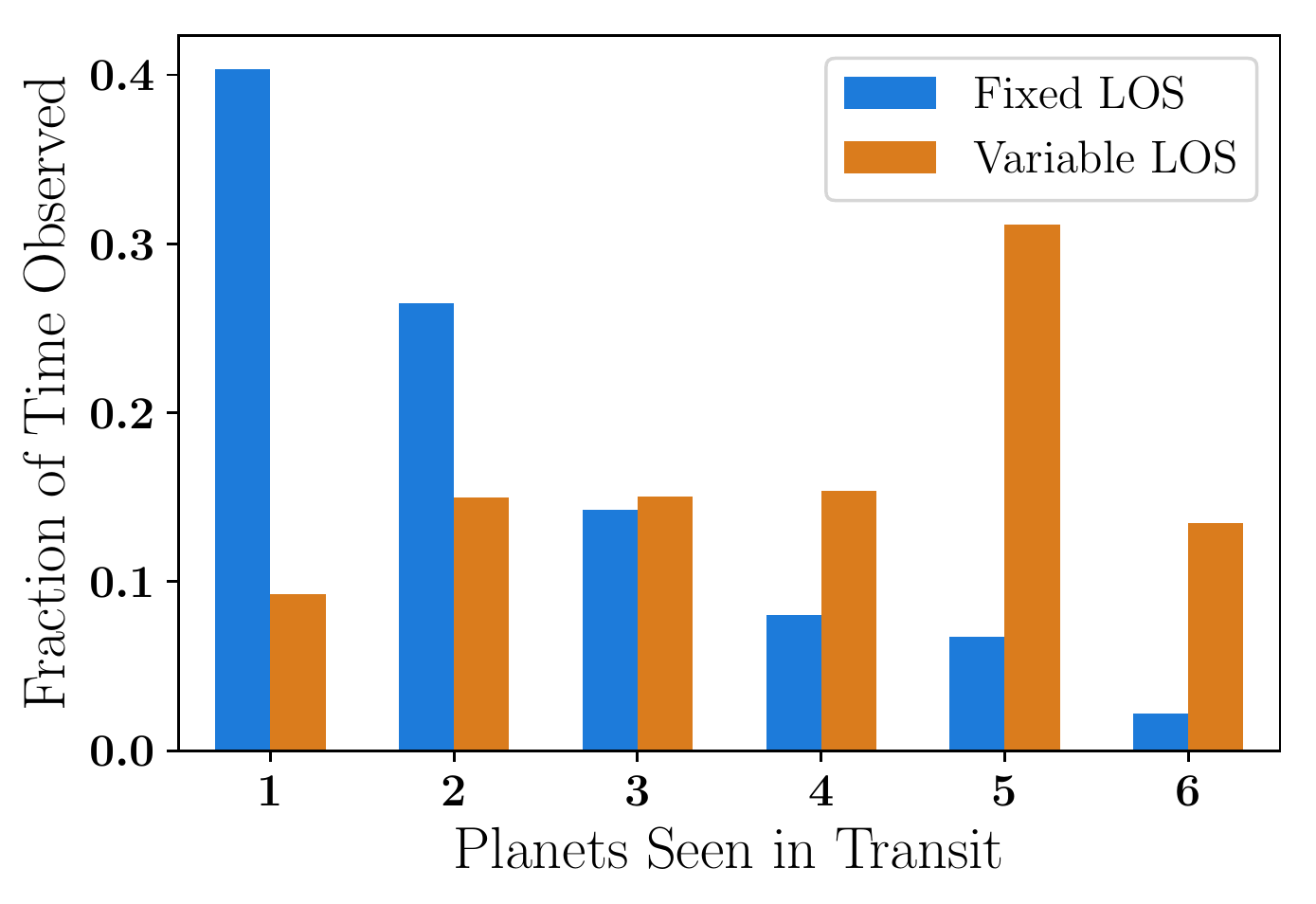}
\caption{The future simultaneous transit fraction by the number of planets seen in transit. These fractions were computed using the numerical simulations for two lines of sight: a fixed line of sight (Fixed LOS), set to be the current line of sight from the earth, and a variable line of sight (Variable LOS), computed at each time-step to be the line of sight from which the greatest number of planets can be seen in transit at any time. The simulations used to construct this plot are the subset of the 250 simulations run in this section. Given the measured orbital elements of the planets around \thisstar, the system is expected to be observed as a six-planet system a minority of the time (assuming there are no extra unseen planets in the system).}
\label{fig:seen_in_transit}
\end{figure}

One notable feature of the numerical simulations is the inclination evolution of all six planets. Secular evolution of the system causes the planetary orbits to evolve with time, resulting in configurations in which not all planets can be seen in transit simultaneously.  Inspired by the present-day (apparently serendipitous) geometry, we extracted from the simulations the transit probability over time for varying numbers of planets in this system. The result of this analysis is presented in Figure \ref{fig:seen_in_transit} for two lines of sight. The first case considers the fixed line of sight corresponding to our current location (that of the Solar System). The second case uses an optimized variable line of sight, which is re-computed at each time-step of each integration to determine the largest multiplicity that can be observed from {\it any} location in the galaxy. 

This analysis shows that observing six transiting planets in the system is rare given the known components of the system. 
No matter which line of sight is considered, the system will appear to contain the six `known' planets a minority of the time. 
More commonly, the system will be seen as a five-planet system from the most favorable line of sight, and as a one- or two-planet system from our current line of sight. 
Most of the time, the inclinations of the outer five planets evolve and cause them to reside in non-transiting configurations. 
On the other hand, the probabilities are not vanishingly small. We expect to be able to re-observe a six-planet transiting system about 2.2 percent of the time in the future from our current line of sight, given the currently measured orbits of these planets.
It is also important to note that we cannot be sure that the observed six planets are the only planets in this system: additional, non-transiting planets would alter the dynamics described here.

\subsection{Resonant state of the two sub-Neptunes} 

The sub-Neptunes \thisstar d and e, with periods of 19.482 days and 14.697 days, have a period ratio of 1.326, which is 0.59\% away from the $4:3$ mean motion commensurability. These planets 
reside in nearly the same orbital plane (with 89.73 and 89.74 degree inclinations). As such, the orbital periods of these planets suggest that they may reside in orbital resonance. However, true resonance is characterized by a librating resonance angle, and it is not clear from the orbital elements alone whether the resonance angle will librate or circulate. To determine the true resonance behavior of these two planets, we used the subset of the 250 simulations which did not experience orbit-crossing during the $10^{5}$ year integation time, and post-process the results to search for resonances. 

We have performed a resonance-finding algorithm to identify the time intervals in the simulations where the planets were in true resonance. We found regimes with nearly constant period ratio (constant $P_e / P_d$), generated arrays of resonance angles for all $p:q$ resonances up to 29th order (while $p \leq 30$), and automatically generated plots using the simulated orbital elements of planets \thisstar d and  e for each resonance angle for each resonance order for all of the 250 simulations. 
Using the resulting resonance angles, we searched for librating behavior by breaking the time series into 5000 year intervals and searching for gaps in resonance angle space: Note that a circulating resonance angle will populate the entire 360 degree range of possible angles, whereas a librating angle will have gaps. 

We find that in our simulations, planets \thisstar d and e exhibit orbital resonances approximately 8.1\% of the time over the simulations under consideration. The resonance angles populated in these cases have the forms 
\begin{equation}
\psi_{1} = 4\lambda_{o} - 3 \lambda_{i} - \varpi_o\,,
\end{equation}
and
\begin{equation}
\psi_{2} = 4\lambda_{o} - 3 \lambda_{i} - \varpi_i \,,\end{equation}
where $\lambda$ is mean longitude and $\varpi$ is longitude of pericenter. The subscripts denote the inner ($i$) and outer ($o$) planets. 
The four types of resonance behavior exhibited by this system, in order of occurrence rate, include: non-resonance, continuous resonance for the entire simulation lifetime, an initial condition close to resonance that loses the resonance as the system evolves, and very rarely, the attainment of a resonance after an initial period of non-resonance (see \citealt{Ketchum:2013} for a more detailed discussion of this process). 
We find that for trials that start out in a resonant configuration, the typical libration width of the resonance is generally consistently around 190 degrees. Although this width may evolve slightly as the simulation progresses, the resonances are not typically much deeper than this initial value.

\subsection{Chaotic Behavior} 

Dynamical systems are often chaotic and we would like to quantify the chaotic behavior of \thisstar. The system, as observed, has six planets in a compact configuration with a relatively large mutual inclination between the innermost planet and the others. Our numerical simulations, described above, indicate that while the outer planets (\thisstar d,  e, and .06) are generally dynamically stable, the middle planets (\thisstar.02 and d) can experience scattering or other non-periodic time evolution, potentially leading to orbit crossing. 
Non-periodic behavior of this nature can be indicative of chaos. 

The evolution of a chaotic planetary system is extremely sensitive to its initial conditions. Chaos is often parameterized by the Lyapunov exponents of the system, which determine the rate of exponential divergence of orbits with similar initial conditions. In contrast, general observational uncertainties in the orbital elements can also lead to non-chaotic divergence if initial orbital elements are drawn from different locations of the posteriors (in cases where ensembles of simulations are used to sample the uncertainties). Either sufficiently large observational errors or the sufficiently rapid onset of chaos will thus make both numerical integrations and analytic explorations less certain. 

\begin{figure}[!ht] 
\centering 
\includegraphics[width=\columnwidth]{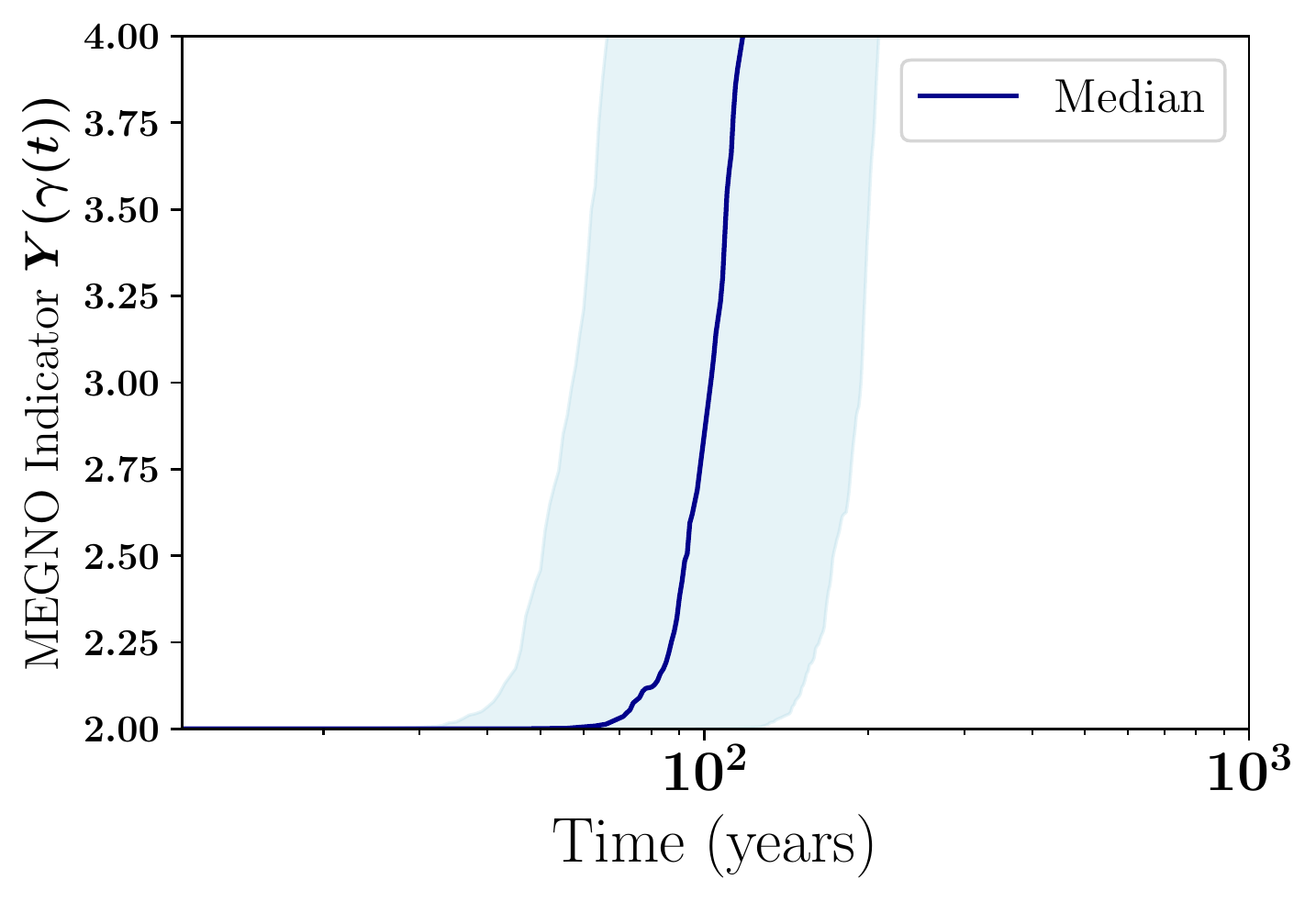}
\caption{The median MEGNO indicator value as a function of time during the thousand-year \texttt{Rebound} integrations (among 400 realizations). The median value reaches $Y=4$, indicating chaotic behavior, after only 100 years. The light blue shaded region delineates the quartile values of the MEGNO indicator. For the majority of posterior draws, this system is thus highly chaotic. }
\label{fig:megno}
\end{figure}

To test the chaotic behavior of the \thisstar system, we ran 400 integrations of this system, with the orbital parameters and masses drawn from the posteriors generated by the transit fit. Each simulation was carried out using the Rebound N-Body integration package \citep{Rein:2012}, where we used a total integration time of 1000 years, the \texttt{IAS15} integrator \citep{Rein:2015}, and an initial time-step of 8 minutes.
For each of these integrations, we evaluate the chaotic nature of the initial conditions by employing the Mean Exponential Growth factor of Nearby Orbits (MEGNO) indicator \citep{Cincotta:2003}, implemented in the \texttt{Rebound} N-body code. 
For chaotic trajectories, the MEGNO indicator, $Y$, grows linearly in time at a rate of $1/t_{Ly}$ where $t_{Ly}$ is the Lyapunov time, while for regular trajectories it asymptotically approaches $Y=2$.
We compute MEGNO values for the 400 draws from the posteriors of this system at times between 1 and 1000 years. These realizations provide a good sample of the parameter space spanned by the observational posteriors. 
Of the 400 realizations, only 4.5\% can be categorized as unambiguously regular at the end of the integration (where the criterion for regularity is taken to be MEGNO$<4$).
Moreover, we find no strong correlations between planetary parameters and MEGNO values using our current simulation set. We attempted to trace chaotic behavior using the period ratio of the resonant planets d and e (as done in Figure 3 of \citealt{Deck:2012}), the ratio perihelion/aphelion of .02 and c, and by using the orbital elements of planet b, but no trends emerged. This finding is likely due to the high multiplicity and tightly packed nature of the system: there are multiple equally-important sources of dynamical chaos. 
In Figure \ref{fig:megno}, we plot the median MEGNO indicator value for these 400 simulations considered in this section at periodic intervals in the 1000 year integrations. A MEGNO indicator less than 4 denotes regular orbits, while a MEGNO of 4 or greater denotes measured chaos. The median MEGNO indicator reaches 4 (denoting the measurable onset of chaos) at roughly 100 years. 

From this MEGNO analysis, we know that the majority of orbits allowed by the transit posteriors are chaotic. It is important to note that, while some of these chaotic posterior points are likely destined to experience instabilities based on our $10^5$ year numerical simulations, a chaotic system does not necessarily mean a dynamically unstable system, or even a particularly active system (the planetary orbits in our solar system are known to be chaotic, as is the Kepler-36 planetary system, \citealt{Deck:2012}). Chaos implies that similar initial conditions will diverge over some time scale, so that precise future predictions of planetary orbits can no longer be made. Specifically, for two given sets of similar initial conditions, integrations of both cases could result in systems that are dynamically stable and continuously transit, but the values of the phase space variables (including planet locations) can diverge over time if the system is chaotic. One implication of this analysis is that a large amount of uncertainty in forward integrations comes from chaos, rather than only from the uncertainty of the transit posteriors.

\section{\thisstarcomp: a likely co-moving companion with a transiting planet candidate }
\label{sec:companion}
After the identification of the 6 possible planets orbiting \thisstar, we searched for nearby stellar companions to understand the full architecture of the system since that may help disentangle the origin of the misaligned inner planet. About $\sim$42$\arcsec$ from \thisstar is \thisstarcomp, which is $\sim$1.7 mag fainter in the $V$-band ($\sim$0.6 mag fainter in the $K$-band).  

\subsection{Evidence for Companionship}
To check that \thisstarcomp is actually an associated companion to \thisstar, we directly compare the Gaia Data Release 2 proper motions and distances for both systems \citep{Gaia:2016, Gaia:2018}. \thisstar has a Gaia parallax of 12.87 $\pm$ 0.06 mas, corresponding to a distance of 77.8$\pm$0.6 pc while \thisstarcomp has a parallax of 12.85 $\pm$ 0.06 mas and a distance of 77.7$\pm$0.6 pc. All systematic uncertainties on the Gaia DR2 parallax should be $<$0.1 mas. Therefore, the two stars are at the same distance. The Gaia DR2 proper motions for \thisstar are $\mu_{\alpha}$ = 56.9 mas yr$^{-1}$ and $\mu_{\delta}$ = -68.8 mas yr$^{-1}$. These are very similar to the proper motions for \thisstarcomp which are $\mu_{\alpha}$ = 53.2 mas yr$^{-1}$ and $\mu_{\delta}$ = -72.7 mas yr$^{-1}$ (see Table \ref{tab:LitProps}). The Gaia parallaxes and proper motions are consistent with the two stars being a widely separated binary.

Using this distance, the $\sim$42$\arcsec$ separation would correspond to a orbital semi-major axis of $\sim$3200 au. Using our determined mass of \thisstarcomp of 0.561\msun\ and \thisstar of 0.649\msun\, this would correspond to an orbital period 0.16 Myr. This period would result in a $\sim$4 mas motion from the binary orbit, within the expected micro-arcsecond expected astrometric precision \citep{Gaia:2018}. However, \thisstarcomp will move $\sim$0.45$\arcsec$ over the nominal 5 year Gaia mission, and it may be difficult to differentiate the contribution from the binary orbit from the star's proper motion. Moreover, as we show in the following section, \thisstarcomp is likely a binary itself, which could further confuse the astrometric solution.

\begin{figure}[!ht]
\vspace{0.3in}
\centering\includegraphics[width=0.99\linewidth, trim = 0.0 0.9in 0.0 1in]{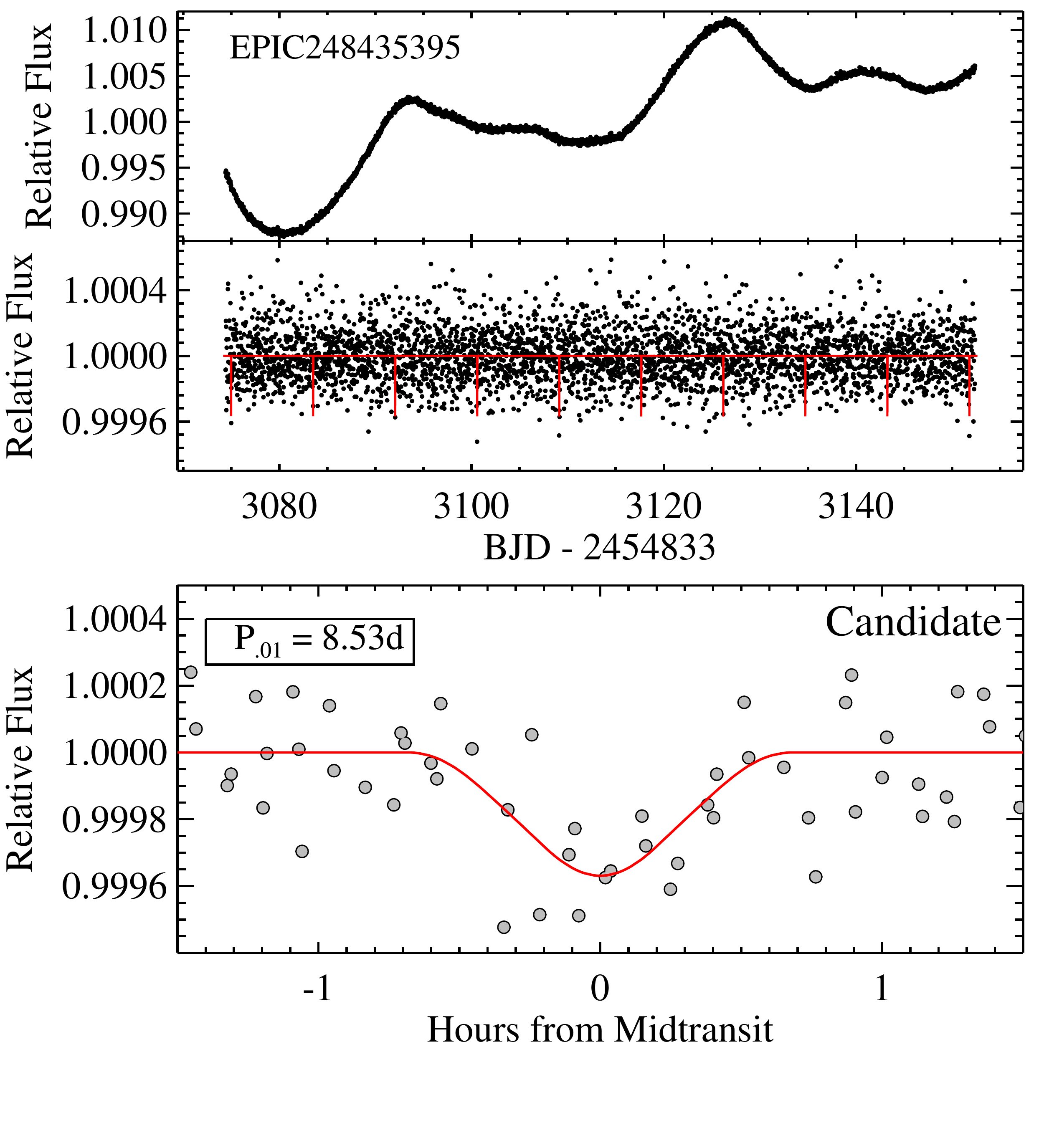}
\caption{(Top) The full K2 light curve of \thisstarcomp from Campaign 14, corrected for systematics using the technique described in \citet{Vanderburg:2014} and \citet{Vanderburg:2016b}. The observations are plotted in open black circles. (Middle) The flattened final {\it K2} light curve  used in the EXOFASTv2 fit. (Bottom) The corrected {\it K2} light curve  phase-folded to the 8.53 day period identified for the planet candidate around \thisstarcomp. The red line corresponds to the final best-fit EXOFASTv2 transit model.  }
\label{figure:LCcompanion}
\end{figure}

\subsection{Observations of \thisstarcomp}
The $\sim$42$\arcsec$ separation means that the two systems were well-resolved by K2. We were therefore able to produce a  separate light curve for the companion star. As we did for \thisstar and following the strategy described in \citet{Vanderburg:2014} and \citet{Vanderburg:2016b}, we searched \thisstarcomp for possible transit signals. From this search, we identify a transit signal at 8.5 days with a S/N of 7.0 around \thisstarcomp, below the typical Kepler S/N threshold of 7.1 and after \Ktwo\ of 9. The final light curve  for \thisstarcomp has a 30 minute noise level of 151 ppm and a 6 hour photometric precision of 42 ppm.

\begin{table}
 \scriptsize
\centering
\setlength\tabcolsep{1.5pt}
\caption{Median values and 68\% confidence intervals for planetary parameters of \thisstarcomp from EXOFASTv2 }
  \begin{tabular}{lccccccccccc}
  \hline
  \hline
\startdata
\multicolumn{2}{l}{Planetary Parameters:}&\thisstarcomp.01\smallskip\\
~~~~$P$\dotfill &Period (days)\dotfill &$8.53467^{+0.00073}_{-0.00063}$\\
~~~~$R_P$\dotfill &Radius (\re)\dotfill &$10.6^{+2.7}_{-8.5}$\\
~~~~$T_C$\dotfill &Time of conjunction (\bjdtdb)\dotfill &$2457907.9884^{+0.0034}_{-0.0039}$\\
~~~~$T_0$\dotfill &Optimal conjunction Time (\bjdtdb)\dotfill &$2457925.0578^{+0.0025}_{-0.0028}$\\
~~~~$a$\dotfill &Semi-major axis (AU)\dotfill &$0.0684\pm0.0012$\\
~~~~$i$\dotfill &Inclination (Degrees)\dotfill &$87.22^{+0.30}_{-0.23}$\\
~~~~$T_{eq}$\dotfill &Equilibrium temperature (K)\dotfill &$530.3^{+7.6}_{-7.7}$\\
~~~~$R_P/R_*$\dotfill &Radius of planet in stellar radii \dotfill &$0.147^{+0.042}_{-0.12}$\\
~~~~$a/R_*$\dotfill &Semi-major axis in stellar radii \dotfill &$22.7^{+1.3}_{-1.4}$\\
~~~~$\delta$\dotfill &Transit depth (fraction)\dotfill &$0.022^{+0.014}_{-0.021}$\\
~~~~$Depth$\dotfill &Flux decrement at mid transit \dotfill &$0.00096^{+0.00020}_{-0.00026}$\\
~~~~$\tau$\dotfill &Ingress/egress transit duration (days)\dotfill &$0.0154\pm0.0016$\\
~~~~$T_{14}$\dotfill &Total transit duration (days)\dotfill &$0.0317^{+0.0061}_{-0.0026}$\\
~~~~$T_{FWHM}$\dotfill &FWHM transit duration (days)\dotfill &$0.0158^{+0.0032}_{-0.0013}$\\
~~~~$b$\dotfill &Transit Impact parameter \dotfill &$1.119^{+0.043}_{-0.14}$\\
~~~~$\delta_{S,3.6\mu m}$\dotfill &Blackbody eclipse depth at 3.6$\mu$m (ppm)\dotfill &$21^{+15}_{-20}$\\
~~~~$\delta_{S,4.5\mu m}$\dotfill &Blackbody eclipse depth at 4.5$\mu$m (ppm)\dotfill &$74^{+49}_{-71}$\\
~~~~$\fave$\dotfill &Incident Flux (\fluxcgs)\dotfill &$0.0179\pm0.0010$\\
~~~~$T_P$\dotfill &Time of Periastron (\bjdtdb)\dotfill &$2457907.9884^{+0.0034}_{-0.0039}$\\
~~~~$T_S$\dotfill &Time of eclipse (\bjdtdb)\dotfill &$2457912.2558^{+0.0031}_{-0.0036}$\\
~~~~$T_A$\dotfill &Time of Ascending Node (\bjdtdb)\dotfill &$2457905.8548^{+0.0035}_{-0.0040}$\\
~~~~$T_D$\dotfill &Time of Descending Node (\bjdtdb)\dotfill &$2457910.1221^{+0.0032}_{-0.0037}$\\
~~~~$d/R_*$\dotfill &Separation at mid transit \dotfill &$22.7^{+1.3}_{-1.4}$\\
~~~~$P_T$\dotfill &A priori non-grazing transit prob \dotfill &$0.0387^{+0.0043}_{-0.0039}$\\
~~~~$P_{T,G}$\dotfill &A priori transit prob \dotfill &$0.0499^{+0.0039}_{-0.0045}$\\
\smallskip\\\multicolumn{2}{l}{Wavelength Parameters:}&Kepler\smallskip\\
~~~~$u_{1}$\dotfill &linear limb-darkening coeff \dotfill &$0.35^{+0.16}_{-0.14}$\\
~~~~$u_{2}$\dotfill &quadratic limb-darkening coeff \dotfill &$0.37^{+0.11}_{-0.14}$\\
\smallskip\\\multicolumn{2}{l}{Transit Parameters:}&dat UT p248-43-53 (Kepler)\smallskip\\
~~~~$\sigma^{2}$\dotfill &Added Variance \dotfill &$0.00000000090^{+0.00000000057}_{-0.00000000054}$\\
~~~~$F_0$\dotfill &Baseline flux \dotfill &$1.0000000\pm0.0000026$\\
\hline
  \hline
\label{tab:ep248435395_1}
 \end{tabular}
\end{table}

We also inspected the NGS POSS and Pan-Stars images of \thisstarcomp\ in a similar manner as described in \S\ref{sec:ArchivalImaging} (see Figure \ref{figure:AO}). Because it shares a common proper motion with \thisstar, it has also moved about 6 arcseconds since it was imaged by POSS in 1952. Because \thisstarcomp\ is fainter than \thisstar, its saturated point spread function does not extend all the way to the star's current-day position, so we are able to rule out background objects down to the POSS limiting magnitude of about 20 in the red, and 19 in the blue. We identified no nearby companions in Pan-STARRS imaging. We also obtained high resolution images in the Br-$\gamma$ filter and the $J$-band of \thisstarcomp using NIRC2 on UT 2017 Dec 28 (see \S\ref{sec:AO} and Figure \ref{figure:AO}). We see no evidence for any additional companions from our ``patient'' and AO imaging. 

We also obtained two spectra of \thisstarcomp using TRES, on UT 2018 Feb 7 and UT 2018 Apr 3. They were analyzed according to the same procedures described in \S\ref{sec:TRES}. The absolute RV of \thisstarcomp is $12.84 \pm 0.31$ \kms, about 2 \kms\ different from \thisstar. Given their projected separation, this is slightly larger than one would expect if the two stars are bound. However, we note that this is unlikely to be the true systemic velocity of \thisstarcomp. The large error on the velocity is not because of poor RV precision, but because our two RVs exhibit a 450 \ms\ variation; \thisstarcomp is likely to be a binary itself, and we do not know the full amplitude of variation or its orbital phase. We conclude that the RVs of the components of the wide pair are not inconsistent with expectations for a bound system; \thisstar is likely to be the primary star in a heierarchical triple system. The TRES RVs for \thisstarcomp are shown in Table \ref{tab:rv}.

\subsection{EXOFASTv2 Global Fit for \thisstarcomp}
Following a similar procedure as we did for \thisstar (see \S\ref{sec:GlobalModel}), we perform a fit of the exoplanet candidate around \thisstarcomp using EXOFASTv2 \citep{Eastman:2017}. Within this analysis, we simultaneously fit the flattened \Ktwo\ light curve  (see Figure \ref{figure:LCcompanion}) and SED (see Figure \ref{fig:sed_fit}) for \thisstarcomp. To constrain the mass of \thisstarcomp within the fit, we use a Gaussian prior of $0.584 \pm 0.015 \msun$ from \citet{Mann:2015}, but with the uncertainties inflated to 5\%. We constrain the radius within the fit using the broad band photometry shown in Table \ref{tab:LitProps}. We set a starting point on the \teff and \feh\ of the host star to be 3699 K and -0.006 dex from the EPIC catalog \citep{Huber:2016}. We also enforce the same upper limit on the $V$-band extinction from the \citet{Schlegel:1998} dust maps of 0.0548 as we did for the fit of \thisstar. We use the Gaia DR2 parallax (12.85 mas) with a conservative uncertainty of 0.1 mas as prior in the fit. The final determined system parameters are shown in Table \ref{tab:ep248435473_0} and \ref{tab:ep248435395_1}. We note that this is not a confirmed or validated planet.  

\section{Discussion}
\label{sec:discussion}

The complex architecture of the planetary system surrounding \thisstar\ makes it an intriguing target for further characterization. Additionally, the host star is relatively bright ($V$=11.8, $K$=8.9) and up to 6 planets orbit the host star in a compact configuration. At the present time, we are only able to validate planets b, c, d, and e, so that more data is needed to confirm the remaining two candidates. 

\subsection{Atmospheric Characterization}

The origin of Neptune sized planets is not clear, yet they appear to be one of the most common types of planets. The planets range in size from 2-6\rearth\ and have been discovered orbiting $>$25\% of all stars \citep{Howard:2012, Fressin:2013, Buchhave:2014, Fulton:2017}. Recent statistical studies of the observed amplitude of transmission spectral features of warm Neptunes show a correlation with equilibrium temperature or its bulk H/He mass fraction \citep{Crossfield:2017b}. However, there are only a small number of Neptune sized planets that are amenable to transmission spectroscopy with current facilities like the {\it Hubble Space Telescope (HST)}. 

To understand if the planets orbiting \thisstar would be viable targets for transmission spectroscopic measurements, we follow the technique described in \citet{Vanderburg:2016b} to calculate their expected atmospheric scale height and signal-to-noise (S/N) per transit. Using data from NASA's Exoplanet archive \citep{Akeson:2013}, we also calculate the atmospheric scale height and S/N for all planets with $R_p$ $<$ 3\rearth (see Table \ref{tbl:S/N}), updating this table from \citet{Rodriguez:2018}. The calculations are done in the $H$-band to understand their accessibility using the Wide Field Camera 3 instrument on HST, as well as the future feasibility to observe them with the suite of instruments on the upcoming James Webb Space Telescope (JWST). Purely based on the inferred sizes of the planets, it is expected that \thisstar\ b, d, and e might all have thick gaseous atmospheres \citep{Weiss:2014}, but the uncertainty in the size of planet b (due to the grazing transit configuration) and its proximity to its host star makes this unclear (see Figure \ref{fig:PDF} for the probability distribution of planet b's radius from our global fit). While our transit fit indicates that the most probable size of planet b is about three times the size of the Earth, virtually all known ultra-short-period planets known are smaller than 2 \rearth\ \citep{Sanchis-Ojeda:2014, Winn:2017}

\thisstar\ b is a particularly interesting target for atmospheric followup because of its status as an ultra-short period (USP) exoplanet. \citet{Lopez:2017} suggests that in order for USP planets to have radii larger than $\sim2 R_\oplus$, they should have formed with high-metallicity, water-rich envelopes, and are likely to remain water-rich today. In addition, the theoretical models of \citet{Owen:2013} and \citet{Owen:2017} suggest that planets with the radial size and orbital period of \thisstar\ b reside near the boundary in parameter space where photoevaporation becomes an important source of mass loss. In addition, if \thisstar b has a typical magnetic field strength, its close proximity allows for interactions between the magnetospheres of the star and the planet \citep{Adams:2011}. Combinations of mass measurements of \thisstar\ b, refined radius measurements, and atmospheric constraints on water abundance could be used together to paint a complete picture of where in the disk this planet originated and when it reached its current-day location. 

\thisstar\ d and e are two of the best sub-Neptune sized planets for atmospheric characterization and their longer periods (as compared to the others in Table \ref{tbl:S/N}) provide a valuable opportunity for a comparative atmospheric study between hot and warm sub-Neptune sized planets. Interestingly, the Near-IR and IR brightness of \thisstar and \thisstarcomp should allow for high S/N observations using short exposure time for all four instruments on JWST: Near Infared Camera (NIRCam), Near Infrared Imager and Slitless Spectrograph (NIRISS), Near-Infrared Spectrograph (NIRSpec), and the Mid-Infrared Instrument (MIRI) \citep{Beichman:2014, Kalirai:2018}.



\subsection{Dynamical Classification}

The dynamics of the \thisstar system is characterized by several remarkable features: The innermost planet (\thisstar b) is highly inclined relative to the rest of the planets, the orbits of planet candidate \thisstar.02 and validated planet \thisstar c are in close proximity, the two sub-Neptunes (\thisstar d, e) are either in or extremely close to a mean-motion resonance, and the outer candidate \thisstar.06 has a moderately eccentric orbit. Taken together, these factors place the planetary system orbiting \thisstar in a particularly unique realm.

Among the exoplanetary systems discovered thus far, only a small number have been determined to be in true resonance \citep{Rivera:2010, Lissauer:2011, Carter:2012, Barclay:2013, Mills:2016,Luger:2017, Shallue:2018, Millholland:2018}. From our numerical simulations initialized with orbital elements from the transit fit, we find in the stable simulations in which no planetary orbits cross, planets \thisstar d and \thisstar e are in true resonance for 8.1\% of the time, as characterized by a librating resonance angle. This significant fraction makes \thisstar another member of the short list of stars hosting systems containing potentially resonant exoplanets. Additional transits, which will improve orbital period precision, would enable future refinement of this resonance fraction. 

From the time-evolution of the MEGNO indicator, we find that the average draw from the posterior becomes noticeably chaotic after roughly 100 years. 
The high mutual inclination between validated planet \thisstar b and the rest of the planets is also intriguing. As shown in Figure \ref{fig:seen_in_transit}, because of the high present day inclination of \thisstar b, this system is rarely (perhaps 5 percent of the time) in a configuration where all six planets can be seen simultaneously from our current line of sight. A smaller number of planets are expected to be seen in transit most of the time. Similarly, it is possible that the system hosts more than six planets, but we are seeing only six of them in transit at the current epoch. Tighter limits on the planetary posteriors will allow for a more precise determination of the future transit probability for each (known) planet, and will place constraints on any additional bodies in the system. 
In future work, a numerical survey of the parameter space subtended by the measured posteriors might allow for additional constraints on planetary parameters based on dynamical stability limits. 

The formation of misaligned orbits, such as that of \thisstar b, remains an open problem. \citet{Petrovich:2018} proposes that most ultra-short period planets form through non-linear secular interactions (``secular chaos''). In this scenario, the proto-USP starts with an orbital period of $5-10$ days, is excited to high eccentricity, and is subsequently tidally captured onto a short-period orbit. Note that the process that leads to high eccentricity (e.g., planet-planet scattering) could also produce inclined orbits. As a result, one potential signature of this process could be a high mutual inclination for the USP relative to the other planets, as seen in this system. However, most companions to a USP generated in this manner would generally have orbital periods of 10 days or larger, whereas the \thisstar system has two planets with shorter periods of only 6 and 7 days. For this secular chaos mechanism to form this system, the 6- and 7-day planets would need to migrate inward after the eccentricity excitement and subsequent circularization of the USP; yet, as \thisstar b may have an un-evaporated atmosphere based on current radius estimates, a dynamical history allowing it to form further from the star (subject to less photoevaporation) seems favorable.

\section{Conclusion}
\label{sec:conclusion}
We present the discovery of up to six planets transiting the bright ($K$=8.9) nearby ($\sim$78 pc) star \thisstar (EPIC248435473). From a global model where we simultaneously fit all six planetary signals, we find that the six planets have periods between 0.66 to 56.7 days, and radii of 0.65 to 3.3 \re. From analyzing transit timing variations, we are able to confirm the two warm Neptunes (d \& e), constraining their masses to be $m_d= 8.9_{-3.8}^{+5.7} M_\oplus$ and $m_e=14.3_{-5.0}^{+6.4} M_\oplus$. Additionally, we are able to validate the planetary nature of planets b and c. Future followup observations should aim to confirm the transits of \thisstar.02 and .06 through high photometric precision observations with facilities like NASA's {\it Spitzer Space Telescope}. Our analysis shows that the inner ultra-short period planet, \thisstar b, has an inclination of 75.3\degr\ while the other five planets are consistent with an inclination between 87--90\degr. This corresponds to a mutual misalignment of $>$12.5\degr\ which may indicate that planet b did not form in the same way as the others. The brightness of \thisstar combined with the relatively large size of the sub-Neptune planets d and e make them great targets for atmospheric characterization observation with current facilities like {\it HST} and future facilities like {\it JWST}.

\acknowledgements

We thank Chelsea Huang, Laura Kreidberg, George Zhou, and Li Zeng for their valuable conversations. Work performed by J.E.R. was supported by the Harvard Future Faculty Leaders Postdoctoral fellowship. AV's contribution to this work was performed under contract with the California Institute of Technology (Caltech)/Jet Propulsion Laboratory (JPL) funded by NASA through the Sagan Fellowship Program executed by the NASA Exoplanet Science Institute.

This work has made use of data from the European Space Agency (ESA) mission {\it Gaia} (\url{https://www.cosmos.esa.int/gaia}), processed by the {\it Gaia} Data Processing and Analysis Consortium (DPAC, \url{https://www.cosmos.esa.int/web/gaia/dpac/consortium}). Funding for the DPAC has been provided by national institutions, in particular the institutions participating in the {\it Gaia} Multilateral Agreement.  

This paper includes data collected by the K2 mission. Funding for the K2 mission is provided by the NASA Science Mission directorate.

A portion of this work was supported by a NASA Keck PI Data Award, administered by the NASA Exoplanet Science Institute. Data presented herein were obtained at the W. M. Keck Observatory from telescope time allocated to the National Aeronautics and Space Administration through the agency's scientific partnership with the California Institute of Technology and the University of California. The Observatory was made possible by the generous financial support of the W. M. Keck Foundation.

The authors wish to recognize and acknowledge the very significant cultural role and reverence that the summit of Maunakea has always had within the indigenous Hawaiian community. We are most fortunate to have the opportunity to conduct observations from this mountain.


\bibliographystyle{apj}

\bibliography{EPIC248435473}

\end{document}